\documentclass[aps,nofootinbib,notitlepage,superscriptaddress]{revtex4-1}
\usepackage{amsmath}
\usepackage{amssymb}
\usepackage{amsfonts}
\usepackage{graphicx}
\usepackage{hyperref}
\usepackage{color}
\usepackage{multirow}
\usepackage{mathtools}
\usepackage{booktabs}
\usepackage{amsthm}
\usepackage{mathtools}
\usepackage{color}
\usepackage{url}

\renewcommand{\thetable}{\arabic{table}}%

\hypersetup{pdftitle=Global financial indices network indicators}
\begin{document}
%-----------------------------------------------------------------
% Title
\title{Network-centric indicators for fragility in global financial indices}
%-----------------------------------------------------------------

%-----------------------------------------------------------------
% Authors
\author{Areejit Samal}
\affiliation{The Institute of Mathematical Sciences (IMSc), 
Homi Bhabha National Institute (HBNI), Chennai 600113 India}
\author{Sunil Kumar}
\affiliation{Department of Physics, Ramjas College, 
University of Delhi, New Delhi 110007 India}
\author{Yasharth Yadav}
\affiliation{Indian Institute of Science Education and Research (IISER), 
Pune 411008 India}
\author{Anirban Chakraborti}
\affiliation{School of Computational and Integrative Sciences,
Jawaharlal Nehru University, New Delhi 110067, India}
\affiliation{Centre for Complexity Economics, Applied Spirituality and 
Public Policy (CEASP), Jindal School of Government and Public Policy, 
O.P. Jindal Global University, Sonipat 131001, India}
\affiliation{Centro Internacional de Ciencias, Cuernavaca 62210,
M\'{e}xico}
%-----------------------------------------------------------------

%-----------------------------------------------------------------
% Abstract
\begin{abstract}
Over the last two decades, financial systems have been studied and analysed 
from the perspective of complex networks, where the nodes and edges in the 
network represent the various financial components and the strengths of 
correlations between them. Here, we adopt a similar network-based approach to 
analyse the daily closing prices of 69 global financial market indices across 
65 countries over a period of 2000-2014. We study the correlations among the 
indices by constructing threshold networks superimposed over minimum spanning 
trees at different time frames. We investigate the effect of critical events 
in financial markets (crashes and bubbles) on the interactions among the indices 
by performing both static and dynamic analyses of the correlations. We compare 
and contrast the structures of these networks during periods of crashes and bubbles, 
with respect to the normal periods in the market. In addition, we study the 
temporal evolution of traditional market indicators, various global network 
measures and the recently developed edge-based curvature measures. We show that 
network-centric measures can be extremely useful in monitoring the fragility in 
the global financial market indices. 
\end{abstract}
%-----------------------------------------------------------------

%-----------------------------------------------------------------
\maketitle
%-----------------------------------------------------------------

%--------------------------------------------------------------------------
%%%% Introduction
\section{Introduction}

It is possible to describe a financial market using the framework of complex 
networks such that the nodes in a network represent the financial components 
and an edge between any two components indicates an interaction between them. 
A correlation matrix constructed using the cross-correlations of fluctuations 
in prices can be utilized to identify such interactions. However, a network 
resulting from the correlation matrix contains densely connected structures. 
A growing amount of research is focused on methods devised to extract relevant 
correlations from the correlation matrix and study the topological, hierarchical 
and clustering properties of the resulting networks.  Mantegna et al. 
\cite{Mantegna1999information,Mantegna1999hierarchical} introduced the minimum 
spanning tree (MST) to extract networks from the correlation matrices computed 
from the asset returns. Dynamic asset trees, introduced by Onnela et al. 
\cite{Onnela2003a,Onnela2004}, were analysed to monitor the evolution of financial 
stock markets using the hierarchical clustering properties of such trees. 
Boginski et al. \cite{Boginski2005} constructed threshold networks by extracting 
the edges with correlation values exceeding a chosen threshold and analyzed 
degree distribution, cliques and independent sets on the threshold network.
Tumminello et al. \cite{Tumminello2005} introduced planar maximally filtered graph 
(PMFG) as a tool to extract important edges from the correlation matrix, which 
contains more information than the MST, while also preserving the hierarchical 
structure induced by MST. Triangular loops and four-element cliques in PMFG could 
provide considerable insights into the structure of financial markets.

Network-based analysis has been widely used to study not only particular stock 
market structures but also the complex networks of correlations among different 
financial market indices across the globe. For example, MST  has been used on 
stock markets to detect underlying hierarchical organization 
\cite{Bonanno2003,Onnela2003,Bonanno2004}. Bonanno et al. \cite{Bonanno2000} 
studied the correlations of 51 global financial indices and showed that the 
corresponding MST was clustered according to the geographical locations of the 
indices. In addition, the changes in the topological structure of MST could 
help understand the evolution of financial systems \cite{Nobi2014,Wang2017,Coelho2007}.
MST and threshold networks have been used to analyse the indices during the 
global financial crisis of 2008 \cite{Kumar2012,Junior2015,Lee2018}. It has also 
been shown that geography is one of the major factors which govern the hierarchy 
of the global market \cite{Leon2017,Saeedian2019}.
Also, Ery\v{g}it and Ery\v{g}it \cite{Eryigit2009} 
had investigated the temporal evolution of clustering networks (MST and PMFG) 
of 143 financial indices corresponding to 59 countries across the world from the 
period 1995-2008, and once again found that the clustering in the networks of 
financial indices was according to their geographical locations. From the time 
dependent network and centrality measures they showed that the integration of 
the global financial indices has increased with time. Further, Chen et al. 
\cite{Chen2020} analyzed dynamics of threshold networks of regional and global 
financial markets from the period 2012-2018, proposed a model for the measurement 
of systemic risk based on network topology and then concluded that network-based 
methods provide a more accurate measurement of systemic risk compared to the 
traditional absorption technique. Silva et al. \cite{Silva2016} studied the 
average criticality of countries during different periods in the crisis and found 
that the USA is the most critical country, followed by European countries, Oceanian 
and Asian countries, and finally Latin American countries and Canada . They also 
found a decrease in the network fragility after the global financial crisis. 
It has been also shown that financial crises can be captured using networks of 
volatility spillovers \cite{Baumohl2018,Mensi2018}. Wang et al. \cite{Wang2018} 
constructed and analysed dynamical structure of MSTs and hierarchical trees computed 
from the Pearson correlations as well as partial correlations, among 57 global 
financial markets from the period 2005-2014, and concluded that MST based on partial 
correlations provided more information when compared to MST based on Pearson 
correlations. The market indices from different stock markets across the globe 
comprise assets that are very different -- apart from stocks of the big multinational 
companies that are traded across markets, the stock markets would have little in common, 
and hence would be expected to behave independently. However, all the aforementioned 
studies suggest in contrary.

In this brief research report, we study the evolution of correlation structures 
among 69 global financial indices through the years 2000 to 2014. To ensure that 
we consider only the most relevant correlations, we construct the network by 
creating an MST (which connects all the nodes) and then add extra edges from the 
correlation matrix exceeding a certain threshold, which gives modular structures. 
Our findings corroborate the earlier results of geographical clustering 
\cite{Leon2017,Sharma2019}. We then study the changes occurring in the market by 
analysing the fluctuations in various global network measures and the recently 
developed edge-based geometric measures. Since there are complex interactions that 
occur among groups of three or more nodes, which cannot be described simply by 
pairwise interactions, the higher-order architecture of complex financial systems 
captured by the geometrical measures can help us in the betterment of systemic 
risk estimation and give us an indication of the global market efficiency. 
To the best of our knowledge, the present work is the first investigation of 
discrete Ricci curvatures in networks of global market indices.
Thus, we find that this approach along with all these network measures can be used 
to monitor the fragility of the global financial network and as indicators of crashes 
and bubbles occurring in the markets.  This could in turn relate the health of the 
financial markets with the development or downturn of the global economy, as well as 
gauge the impact of certain market crises in the multi-level financial-economic 
phenomena.

%--------------------------------------------------------------------------

% --------------------------------------------------------------------------
% Methods
\section{Methods}
% --------------------------------------------------------------------------

% --------------------------------------------------------------------------
\subsection{Data description}

This study is based on a dataset collected from Bloomberg which comprises 
the daily closing prices of 69 global financial market indices from 65 countries, 
and this information was compiled for a period of $T=3513$ days over 14 years from 
11 January 2000 to 24 June 2014. Note that the working days for different markets 
are not same due to differences in holidays across countries. To overcome any 
inconsistencies due to this difference in working days, we filtered the data by 
removing days on which $> 30\%$ of the markets were not operative. Conversely, 
if $< 30\%$ of the markets were not operative on a day, we used the closing price 
of such markets on the previous day to complete the dataset. Supplementary Table S1 
lists the 69 global market indices considered here, along with their countries 
and geographical regions.   

% --------------------------------------------------------------------------

% --------------------------------------------------------------------------
\subsection{Cross-correlation matrix and market indicators}

Given the daily closing price $g_j(t)$ for market index $j$ on day 
$t$, wherein $j=1,2,\ldots,N$ with $N=69$ indices, we construct a time series 
of logarithmic returns as $r_j(t) = \ln g_j(t) - \ln g_j(t-1)$. Then, we 
construct the equal time Pearson cross-correlation matrix as
\begin{equation}
\label{eq:corr}
C_{ij}^{\tau}(t) = \frac{\left \langle r_i r_j \right \rangle - \left \langle 
r_i \right \rangle \left \langle r_j \right \rangle}{\sigma_i \sigma_j},
\end{equation}
where the mean and standard deviation are computed over a period of $\tau=80$ 
days with end date as $t$. We also construct the ultrametric distance matrix 
with elements $D_{ij}^{\tau}(t) = \sqrt{2(1- C_{ij}^{\tau}(t))}$ that take 
values between 0 and 2. To study the temporal dynamics of the global market 
indices, we computed the correlation matrices for overlapping windows of 
$\tau=80$ days with a rolling shift of $\Delta\tau=20$ days. Thence, we 
obtained $172$ correlation frames between 11 January 2000 to 24 June 2014.

We have computed three market indicators from these correlation matrices 
$\mathbf{C}^{\tau}(t)$. Firstly, the mean correlation gives the average of 
the correlations in the matrix $\mathbf{C}^{\tau}(t)$. Secondly, we have 
computed the eigen-entropy \cite{Chakraborti2020a} which involves calculation 
of the Shannon entropy using the eigenvector centralities of the correlation 
matrix $\mathbf{C}^{\tau}(t)$ of market indices. Both mean correlation
and eigen-entropy has been shown to detect critical events in financial markets 
\cite{Chakraborti2020a,Chakraborti2020b,Kukreti2020}. Thirdly, we have computed 
the risk corresponding to the Markowitz portfolio of the market indices, 
which is a proxy for the fragility or systemic risk of the global financial 
network \cite{Sandhu2016}. A detailed description of the Markowitz portfolio 
optimization is given in the Supplementary Material.

% --------------------------------------------------------------------------

% --------------------------------------------------------------------------
\subsection{Threshold network construction and characteristics}

The distance matrix for the time frame ending on $t$ can be viewed as a 
complete, undirected and weighted graph $\mathbf{D}^{\tau}(t)$ where 
the element $D_{ij}^{\tau}(t)$ is the weight of the edge between market indices $i$ 
and $j$. To extract the important edges from $\mathbf{D}^{\tau}(t)$, we first 
construct its minimum spanning tree (MST) $\mathbf{M}^{\tau}(t)$ using Prim's 
algorithm \cite{Prim1957}. As MST is an over-simplified network without cycles, 
it may lose crucial information on clusters or cliques. To overcome this, we add edges 
with correlation $C_{ij}^{\tau} \geq 0.65$ in $\mathbf{D}^{\tau}(t)$ to 
$\mathbf{M}^{\tau}(t)$ and obtain the threshold graph $\mathbf{S}^{\tau}(t)$. 
Thereafter, we study the temporal evolution of different network measures in 
$\mathbf{S}^{\tau}(t)$.

Firstly, we have computed standard global network measures such as the number of 
edges, edge density, average degree, average weighted degree \cite{Barrat2004}, 
average shortest path length, diameter, average clustering coefficient 
\cite{Onnela2005}, modularity \cite{Girvan2002,Blondel2008}, communication 
efficiency \cite{Latora2001}, global reaching centrality (GRC) \cite{Mones2012}, 
network entropy \cite{Sole2004}, global assortativity \cite{Newman2003, Leung2007} 
and clique number. 
Note that the chosen set of global network measures studied here are 
by no means exhaustive and also depend very much on the specific questions of 
interest, see for example, Wang et al. \cite{Wang2018b} for several gravitational
centrality measures.
Secondly, we have also computed four edge-centric curvature 
measures, namely, Ollivier-Ricci (OR) curvature \cite{Ollivier2007,Sandhu2016,Samal2018}, 
Forman-Ricci (FR) curvature \cite{Forman2003,Sreejith2016,Samal2018,Saucan2019a}, 
Menger-Ricci (MR) curvature \cite{Saucan2019b,Saucan2020} and Haantjes-Ricci (HR) 
curvature \cite{Saucan2019b,Saucan2020}. A detailed description of these network 
measures along with the appropriate natural weight, strength or distance, to 
use in each case is included in the Supplementary Material.

% --------------------------------------------------------------------------

% --------------------------------------------------------------------------
\subsection{Multidimensional scaling map}

The multidimensional scaling (MDS) technique tries to embed $N$ objects in 
high-dimensional space into a low-dimensional space (typically, $2$- or 
$3$-dimensions), while preserving the relative distance between pairs of 
objects \cite{Borg2005}. Here, we construct the (average) correlation matrix 
$\mathbf{C}^{T}$ between the 69 market indices for the complete period of 
$T=3513$ days between 11 January 2000 to 24 June 2014 using Eq. \ref{eq:corr}. 
Then, we compute the distance matrix $\mathbf{D}^{T}$ from $\mathbf{C}^{T}$ 
for the complete period. Thereafter, we use MDS to map the 69 market indices 
into a $2$-dimensional space such that the distances between pairs of indices 
in $\mathbf{D}^{T}$ are preserved. To create the MDS plot, we used the in-built 
function cmdscale.m in \texttt{MATLAB}. Moreover, we also construct the MST 
$\mathbf{M}^{T}$ starting from the distance matrix $\mathbf{D}^{T}$, and then, 
the threshold network $\mathbf{S}^{T}$ for the complete period from 2000 to 
2014 by adding edges with $C_{ij}^{T} \ge 0.65$ to $\mathbf{M}^{T}$.

% --------------------------------------------------------------------------

%--------------------------------------------------------------------------
% Figure 1
\begin{figure}
\begin{center}
\includegraphics[width = 0.79\linewidth]{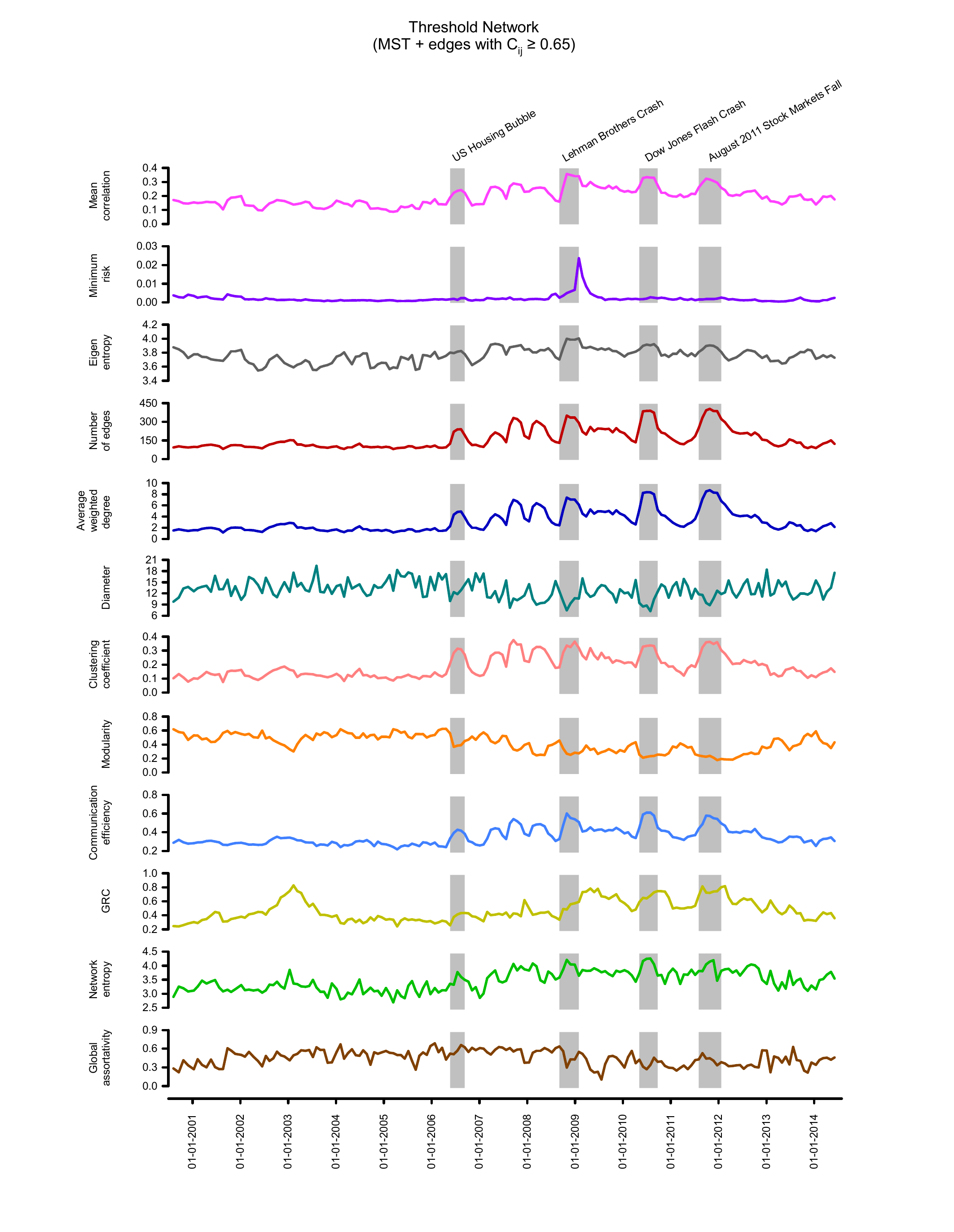}
\end{center}
\caption{Evolution of generic indicators and network characteristics for the 
global market indices networks $\mathbf{S}^{\tau}(t)$, constructed from the 
correlation matrices $\mathbf{C}^{\tau}(t)$ of window size $\tau=80$ days and 
an overlapping shift of $\Delta\tau=20$ days over a period of 14 years 
(2000-2014). The threshold networks $\mathbf{S}^{\tau}(t)$ were constructed 
by adding edges with correlation $C_{ij}^{\tau}(t) \ge 0.65$ to the minimum 
spanning trees (MST). From top to bottom, we compare the plots of mean 
correlation among market indices, minimum risk corresponding to the Markowitz 
portfolio optimization, eigen-entropy, number of edges, average weighted degree, 
diameter, clustering coefficient, modularity, communication efficiency, global 
reaching centrality (GRC), network entropy and global assortativity. The four 
shaded regions correspond to the epochs around the four important market events, 
namely,  US housing bubble, Lehman brothers crash, Dow Jones flash crash, and 
August 2011 stock markets fall.}
\label{Fig1}
\end{figure}
%-------------------------------------------------------------------------

%--------------------------------------------------------------------------
% Figure 2
\begin{figure}
\begin{center}
\includegraphics[width = 0.72\linewidth]{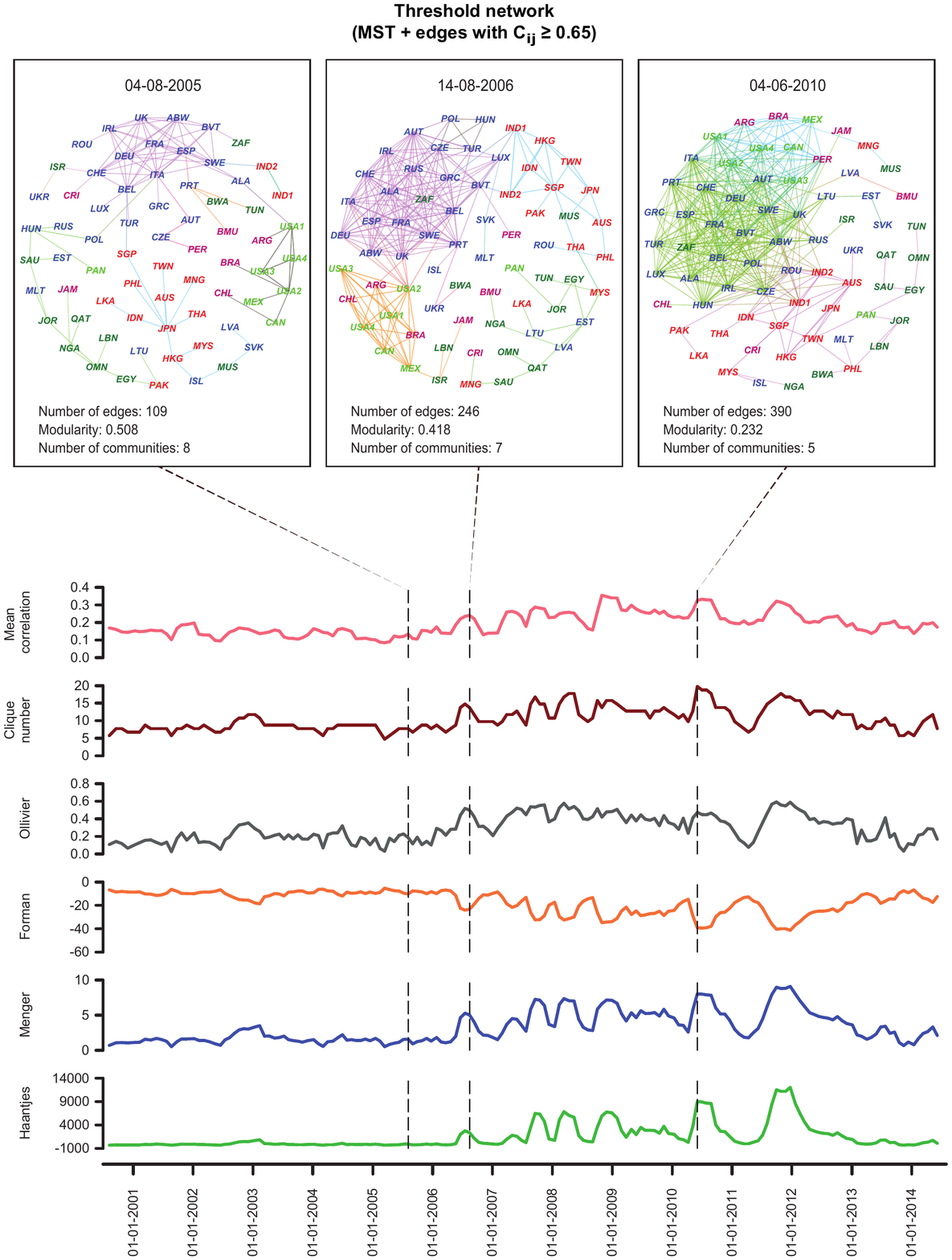}
\end{center}
\caption{Evolution of network characteristics and visualization of the 
threshold networks $\mathbf{S}^{\tau}(t)$ of market indices with window 
size $\tau=80$ days and an overlapping shift of $\Delta\tau=20$ days, 
constructed by adding edges with correlation $C_{ij}^{\tau}(t) \ge 0.65$ 
to the MST. (Lower panel) Comparison of the plots of mean correlation 
among market indices, clique number, average of Ollivier-Ricci (OR), 
Forman-Ricci (FR), Menger-Ricci (MR), and Haantjes-Ricci (HR) curvature 
of edges in threshold networks over the 14-year period. (Upper panel) 
Visualization of the threshold networks at three distinct epochs of 
$\tau=80$ days ending on trading days $t$ equal to 04-08-2005 (normal), 
14-08-2006 (US housing bubble) and 04-06-2010 (Dow Jones flash crash). 
Threshold networks show higher number of edges and lower number of 
communities during crisis. Correspondingly, there is an increase in mean 
correlation, clique number, average OR, MR and HR curvature, and decrease 
in average FR curvature of threshold networks during financial crisis. 
Node colours and labels are based on geographical region and country, 
respectively, of the indices and edge colours are based on the community 
determined by Louvain method. The four USA market indices, NASDAQ, NYSE, 
RUSSELL1000 and SPX, are labelled as USA1, USA2, USA3 and USA4, respectively, 
while the two Indian indices, namely, NIFTY and SENSEX30, are labelled as 
IND1 and IND2, respectively.}
\label{Fig2}
\end{figure}
%-------------------------------------------------------------------------

%--------------------------------------------------------------------------
% Figure 3
\begin{figure}
\begin{center}
\includegraphics[width = 0.7\linewidth]{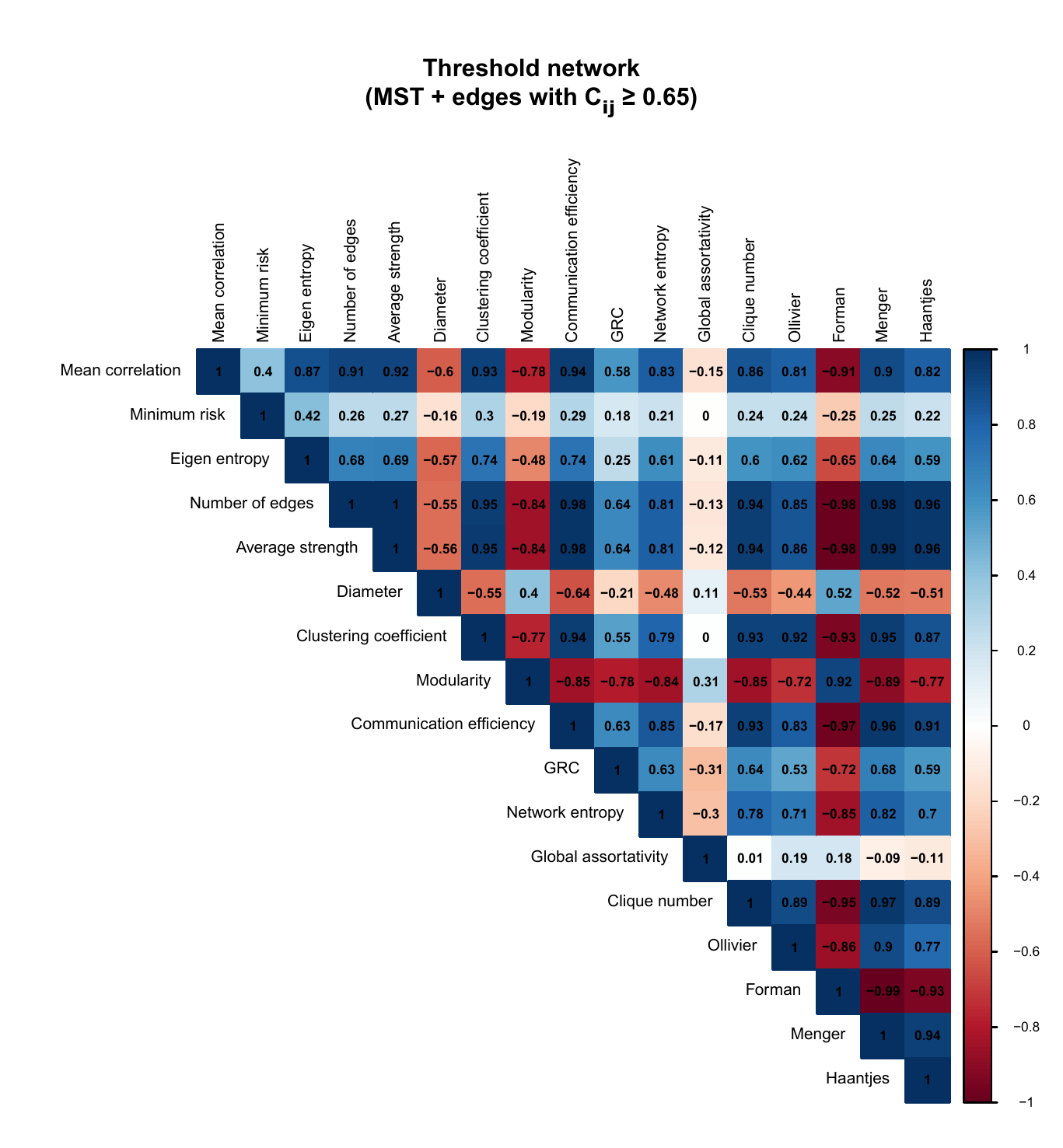}
\end{center}
\caption{Correlations between generic indicators and network characteristics 
of the global market indices networks $\mathbf{S}^{\tau}(t)$, constructed 
from the correlation matrices $\mathbf{C}^{\tau}(t)$ of window size $\tau=80$ 
days and an overlapping shift of $\Delta\tau=20$ days over a period of 14 
years (2000-2014). The threshold networks $\mathbf{S}^{\tau}(t)$ were 
constructed by adding edges with correlation $C_{ij}^{\tau}(t) \ge 0.65$ to 
the minimum spanning tree (MST).}
\label{Fig3}
\end{figure}
%-------------------------------------------------------------------------

%--------------------------------------------------------------------------
% Figure 4
\begin{figure}
\begin{center}
\includegraphics[width = 0.79\linewidth]{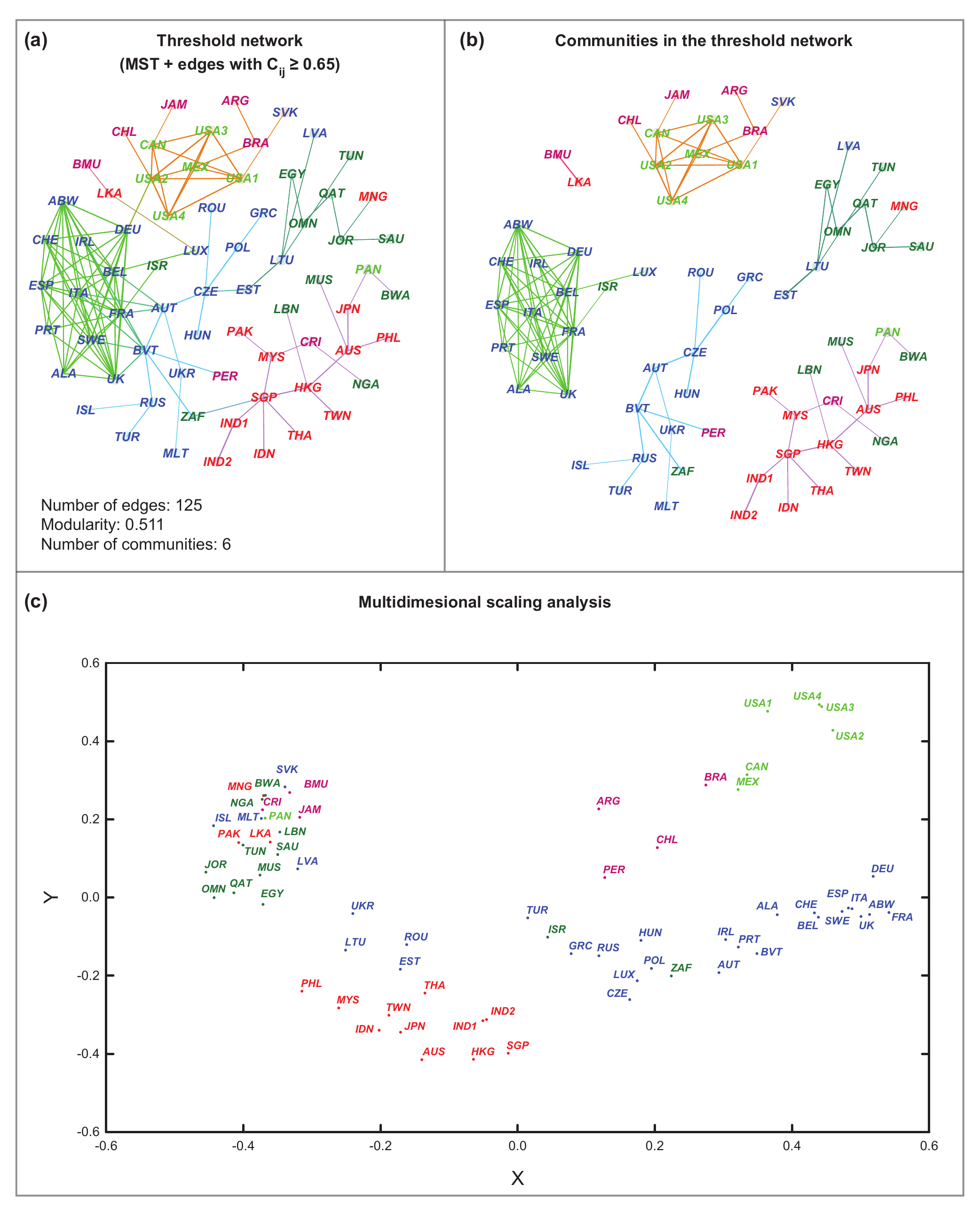}
\end{center}
\caption{The average correlation structure between 69 market indices over the 
14-year period is visualized based on the correlation matrix $\mathbf{C}^{T}$ 
for the complete period of $T=3513$ days between 2000 to 2014. {\bf (a)} Visualization
of the overall threshold network $\mathbf{S}^{T}$ corresponding to $\mathbf{C}^{T}$ 
obtained by combining MST plus edges with correlation $\ge 0.65$. The node colours 
are based on geographical regions of the market indices and edge colours are based 
on communities obtained from Louvain method. {\bf (b)} Visualization of the 
communities in the overall threshold network $\mathbf{S}^{T}$ after removing the 
inter-module edges. It is evident that the market indices form communities in this 
network based on their geographical proximity. {\bf (c)} Multidimensional scaling 
(MDS) map in $2$-dimensions of the 69 market indices. In this figure, the indices 
are labelled in different colours based on their geographical region and country, 
respectively. The four USA market indices, NASDAQ, NYSE, RUSSELL1000 and SPX, are 
labelled as USA1, USA2, USA3 and USA4, respectively, while the two Indian indices, 
NIFTY and SENSEX30, are labelled as IND1 and IND2, respectively.}
\label{Fig4}
\end{figure}
%-------------------------------------------------------------------------

%-------------------------------------------------------------------------    
% Results and Discussion
\section{Results and Discussion}

The primary goal of this investigation is to evaluate different network 
measures for their potential to serve as indicators of fragility or systemic risk and 
monitor the health of the global financial system. For this purpose, we 
compiled a dataset of the daily closing prices of 69 global financial 
market indices from 65 different countries for a 14-year period from 2000 
to 2014 (Methods). Thereafter, we use the time-series of the logarithmic 
returns of the daily closing prices for 69 global market indices to compute 
the Pearson cross-correlation matrices $\mathbf{C}^{\tau}(t)$ with window 
size of $\tau=80$ days with overlapping shift of $\Delta \tau=20$ days, and 
ending on trading days $t$ (Methods). Subsequently, we employ a minimum 
spanning tree (MST) based approach to construct 172 threshold networks 
$\mathbf{S}^{\tau}(t)$ corresponding to the cross-correlation matrices 
$\mathbf{C}^{\tau}(t)$ spanning the 14-year period (Methods). Here, we 
study the temporal evolution of the structure of these correlation-based 
threshold networks $\mathbf{S}^{\tau}(t)$ of global market indices using 
several network measures, and moreover, contrast the evolution of network
properties with generic market indicators such as mean correlation and minimum 
risk obtained using Markowitz framework. 

We reiterate that the threshold networks $\mathbf{S}^{\tau}(t)$ are constructed 
by computing the MST of the cross-correlation matrices $\mathbf{C}^{\tau}(t)$ 
followed by addition of edges with correlation $C_{ij}^{\tau} \geq 0.65$ 
(Methods). Intuitively, this network construction procedure ensures that 
each threshold network is a connected graph and captures the most relevant
edges (correlations) between market indices. Since the obtained results may 
depend on the choice of the threshold ($0.65$) used for network construction, 
we present the temporal evolution of properties in networks constructed using 
$0.65$ as threshold in Main text, and in networks constructed using $0.75$ or 
$0.85$ as threshold in Supplementary Material. In the sequel, we will show 
that the qualitative nature of the obtained results are not very sensitive to the 
choice of $0.65$, $0.75$ or $0.85$ as thresholds to construct the networks of 
global market indices.  

In Figures \ref{Fig1}, \ref{Fig2} and Supplementary Figure S1, we show the 
temporal evolution of generic indicators and network measures in the threshold
networks of global market indices over the 14-year period (2000-2014). Moreover, 
the four shaded regions in Figure \ref{Fig1} highlight four periods of financial 
crisis, namely, US housing bubble, Lehman brothers crash, Dow Jones flash crash, 
and August 2011 stock markets fall. From Figure \ref{Fig1}, it is seen that the 
mean correlation between market indices increases during periods of financial 
crisis. Also, the eigen-entropy which is directly computed from the correlation 
matrix $\mathbf{C}^{\tau}(t)$ increases during crisis. Earlier works have shown 
that mean correlation and eigen-entropy are indicators of instabilities in stock 
market network \cite{Chakraborti2020a,Kukreti2020}, and we show here that these measures can 
also serve as indicators of crisis in network of global financial indices. In 
Figure \ref{Fig1}, we also show the temporal evolution of the minimum risk 
corresponding to the portfolio comprising the market indices using the Markowitz 
framework. Moving on to widely-used network properties, it is seen that the number 
of edges, edge density, average degree, average weighted degree, clustering 
coefficient, communication efficiency and network entropy increase while diameter, 
average shortest path length and modularity decrease during periods of financial 
crisis (Figure \ref{Fig1}; Supplementary Figure S1). In Figure \ref{Fig1}, we also 
show evolution of two other network measures, global reaching centrality (GRC) and 
global assortativity. In Figure \ref{Fig2}, we also visualize the threshold network 
at three distinct time windows of $\tau=80$ days ending on trading days $t$ 
corresponding to 04-08-2005 (normal period), 14-08-2006 (US housing bubble 
crisis) and 04-06-2010 (Dow Jones flash crash) where the node colours are based on 
geographical regions of the market indices and edge colours are based on modules 
determined by Louvain method \cite{Blondel2008} for community detection. The 
identified communities in the three networks corresponding to normal period, US 
housing bubble and Dow Jones flash crash typically reflect the geographical 
proximity of financial market indices. For example, the indices of USA, Canada,  
Mexico, Argentina, Brazil and Chile form a single community in the threshold network
for the normal period (Figure \ref{Fig2}). It is evident that the number of edges 
in threshold networks correspoding to US housing bubble (246 edges) or Dow Jones 
flash crash (390 edges) are much higher in comparison to that for normal period 
(109 edges). In contrast, the modularity of threshold networks corresponding to 
the crisis periods, US housing bubble (0.418) or Dow Jones flash crash (0.232) are 
lower in comparison to that for normal period (0.508). In Figure \ref{Fig2}, 
it is clearly seen that the clique number or size of the largest clique in threshold 
networks increases during financial crisis, and this is also evident from the network 
visualizations for normal period, US housing bubble and Dow Jones flash crash. 
Note that bubbles are not easy to detect. In fact, our proposition is that 
holistic approaches with network measures, both node- and edge-based measures, 
including geometric curvatures, may help us to better detect and distinguish the 
bubbles from market crashes, as also pointed out in recent contributions 
\cite{Chakraborti2020a,Samal2020}.
In sum, we find that during a normal period the network of global market indices is less 
connected, very modular and heterogeneous, whereas during a fragile period the network 
is highly connected, less modular and more homogeneous. 

In addition to the node-centric global network measures described in the preceding 
paragraph, we have also studied edge-centric network measures, specifically, four 
discrete Ricci curvatures [Olivier-Ricci (OR), Forman-Ricci (OR), Menger-Ricci (MR) 
and Haantjes-Ricci (HR)] in threshold networks of global market indices. From Figure 
\ref{Fig2}, it is seen that the average OR, MR or HR curvature of edges increase during 
crisis periods in comparison to normal periods. In contrast, the average FR curvature 
of edges decreases during crisis periods in comparison to normal periods. Notably, 
Sandhu et al. \cite{Sandhu2016} have shown that OR curvature can serve as 
indicator of fragility in stock market networks. However, to our knowledge, the 
present work is the first investigation of discrete Ricci curvatures 
in networks of global market indices. Note that different discretizations of Ricci 
curvature do not capture the entire features of the classical definition for continuous 
spaces, and thus, the four discrete Ricci curvatures studied here can capture different 
aspects of analyzed networks \cite{Samal2018}. Overall, our results suggest that 
discrete Ricci curvatures can serve as indicators of fragility and monitor the health 
of the global financial system. 

In Figure \ref{Fig3}, we show the correlation between generic market indicators and 
different characteristics of the threshold networks $\mathbf{S}^{\tau}(t)$ of global 
market indices computed across the 14-year period from 2000 to 2014. From this figure, 
it is seen that eigen-entropy and several network measures have a very high (absolute) 
Pearson correlation ($\approx 0.9$) with generic indicator, mean correlation of market 
indices. Such network measures incude number of edges, average weighted degree (strength), 
clustering coefficient, communication efficiency, clique number, FR curvature and MR 
curvature. In contrast to mean correlation of market indices, there is moderate to 
no correlation between minimum risk corresponding to the portfolio comprising the market 
indices and eigen-entropy or network measures (Figure \ref{Fig3}). In sum, these results 
indicate that network measures including edge-centric FR curvature can be used to forecast 
crisis and monitor the health of the global financial system. To the best of our knowledge, 
our work is the largest survey of network measures to identify potential network-centric
indicators of fragility in global financial market indices. 

We must mention that though in the preceding paragraphs we have described only the 
results obtained from networks 
constructed using threshold of $0.65$, we have shown in Supplementary Figures S2-S9 
that the qualitative conclusions remain unchanged even when networks with 
threshold of $0.75$ and $0.85$ are considered. In other words, our results are 
robust to the choice of threshold used to construct the networks of global market indices. 

In previous works, the econophysics community has employed either minimum spanning 
tree (MST) \cite{Bonanno2000,Bonanno2003,Bonanno2004,Coelho2007,Eryigit2009,Nobi2014,
Junior2015,Wang2017} or planar maximally fitered subgraph (PMFG) 
\cite{Eryigit2009,Wang2017} or threshold networks \cite{Kumar2012,Nobi2014,Chen2020} 
to study the correlation structure between global financial market indices. 
As far as we know, this work is the first to use threshold networks of MST plus edges 
with correlation higher than a specified threshold, to study the temporal evolution of 
relationships between global financial market indices. In contrast, such threshold 
networks based on MST have been used earlier to study the structure of stock market
networks \cite{Sandhu2016,Samal2020}. While MST has a tree structure without loops 
or cycles, PMFG or threshold network permit loops or cycles. In Supplementary Text
and Figures S10-S13, we also display the temporal evolution and correlation between 
generic market indicators and network measures in PMFG of global market indices 
constructed from cross-correlation matrices $\mathbf{C}^{\tau}(t)$. While the 
construction of PMFG unlike threshold networks is independent of any specific choice
of the threshold, the number of edges (thus, edge density and average degree) is fixed 
in case of PMFG (Supplemetary Figures S10 and S11). Due to this reason, we find that
most of the network measures studied here are not correlated with the generic market 
indicator, mean correlation of market indices, in PMFG case (Supplementary Figure 
S13). Still, we find that average weighted degree (strength), clustering coefficient 
and communication efficiency have very high correlation with mean correlation of 
market indices in PMFG based networks (Supplementary Figure S13). Based on these results, 
the threshold network construction based on MST plus edges with high correlation seems 
a better framework to monitor the state of the global financial system.

Finally, we have also studied the average correlation structure between global market 
indices over the 14-year period by computing the correlation matrix $\mathbf{C}^{T}$ 
between the 69 market indices by taking window size as the complete period of $T$ days 
between 2000 to 2014 (Methods). Subsequently, we have constructed a threshold network 
$\mathbf{S}^{T}$ corresponding to $\mathbf{C}^{T}$ by combining MST plus edges with 
correlation above the chosen threshold of $0.65$ (Methods). In Figure \ref{Fig4}(a), 
we visualize this overall threshold network $\mathbf{S}^{T}$ of market indices for 
the complete 14-year period of $T$ days. In this figure, the node colours are based on 
geographical regions of the market indices and edge colours are based on communities 
obtained from Louvain method. In Figure \ref{Fig4}(b), we have separated the communities 
in this overall threshold network $\mathbf{S}^{T}$ of market indices by removing the 
inter-module edges in the visualization. From Figure \ref{Fig4}(a,b), it is clear that 
the market indices form communities in this overall threshold network based on their  
geographical proximity. Moreover, we have also employed multidimensional scaling (MDS) 
technique to map the 69 market indices into a $2$-dimensional space such that the 
distances between pairs of indices are preserved (Figure \ref{Fig4}(c); Methods). It 
can be seen that the MDS map is able to partition the 69 market indices into groups based 
on their geographical proximity, and further, the structure in the MDS map has close
resemblance to the community structure of the overall threshold network (Figure \ref{Fig4}).
For example, the grouping of indices from USA, Canada, Mexico, Argentina, Brazil and 
Chile can be seen in both the threshold network and MDS map (Figure \ref{Fig4}).   
Interestingly, when we plotted in Supplementary Figure S14, the evolution of the eigenvector 
centralities of the nodes (market indices), as well as their OR and FR curvature, we 
found that there exist certain periods of time, when some of the countries in close 
geographical proximity display high (absolute) values and others display low values, 
indicative of the changes in the complex interactions and community structures.    

%-------------------------------------------------------------------------

% --------------------------------------------------------------------------
% Conclusions
\section{Summary and concluding remarks}

In summary, we have investigated the daily closing prices of 69 global financial 
indices over a 14-year period using various techniques of cross-correlations 
based network analysis. We have been able to continuously monitor the complex 
interactions among the global market indices by using a variety of network-centric 
measures, including, recently developed edge-centric discrete Ricci curvatures. 
In the present study of the global market indices, the novelty lies in: 
(i) Construction of the threshold network $\mathbf{S}^{\tau}(t)$, as superposition 
of the MST of the cross-correlation matrix and the network of edges with correlations 
$C_{ij}^{\tau} \geq 0.65$, which ensures that each threshold network is a connected 
graph and captures the most relevant edges (correlations) between market indices. 
In Supplementary Material, we have also reported the results for networks constructed 
using MST and two other threshold values, i.e., $C_{ij}^{\tau} \geq 0.75$ and 
$C_{ij}^{\tau} \geq 0.85$. Besides, we have also reported results for networks 
constructed using PMFG method. 
(ii) The usage of discrete Ricci curvatures in networks of global market indices, 
which capture the higher-order architecture of the complex financial system. 
To the best of our knowledge, this is the first study employing edge-based 
discrete Ricci curvatures to networks of global financial indices. Our recent work
underscores the utility of edge-based curvature measures in analysis of networks of 
stocks \cite{Samal2020} or global financial indices. In future, curvature measures may 
also find application in other financial networks including Banking networks 
\cite{Namaki2020}.
(iii) The largest yet by no means exhaustive
survey of network measures to identify potential network-centric
indicators of fragility and systemic risk in the system of global financial market 
indices. 

The global financial system has become increasingly complex and interdependent, 
and thus prone to sudden unpredictable changes like market crises.  Our results, 
compared to the traditional market indicators, do provide a deeper understanding 
of the system of global financial markets. Specially, we find that the four 
discrete Ricci curvatures can be effectively used as indicators of fragility 
in global financial 
markets. We reiterate that the methods used in this work can detect instabilities 
in the market, and can be used as early warning signals so that policies can be 
made in order to prevent the occurrence of such events in the future. 

% --------------------------------------------------------------------------

% --------------------------------------------------------------------------

% --------------------------------------------------------------------------
% Data Availability Statement
\section*{Data Availability Statement}
The codes used to construct the networks from correlation matrices and compute 
the different network measures are publicly available via the GitHub repository: 
\url{https://github.com/asamallab/FinNetIndicators}.  

% --------------------------------------------------------------------------

% --------------------------------------------------------------------------
% Author Contributions
\section*{Author Contributions}

A.S., S.K. and A.C. conceived the project. A.S., S.K., Y.Y. and A.C. 
performed the computations. S.K. compiled the dataset. Y.Y. and A.S. 
prepared the figures and tables. A.S. and A.C. analyzed the results. 
A.S., S.K., Y.Y. and A.C. wrote the manuscript. All authors have read 
and approved the manuscript. 
% --------------------------------------------------------------------------

% --------------------------------------------------------------------------
% Conflict of Interest
\section*{Conflict of Interest}

The authors declare that the research was conducted in the absence of any 
commercial or financial relationships that could be construed as a potential 
conflict of interest.
% --------------------------------------------------------------------------

% --------------------------------------------------------------------------
% Acknowledgments
\section*{Acknowledgments}

A.C. acknowledges support from the project UNAM-DGAPA-PAPIIT AG 100819 and 
CONACyT Project FRONTERAS 201. A.S. acknowledges financial support from Max 
Planck Society Germany through the award of a Max Planck Partner Group in 
Mathematical Biology and a Ramanujan fellowship (SB/S2/RJN-006/2014) from 
the Science and Engineering Research Board (SERB), India.

% --------------------------------------------------------------------------

%-----------------------------------------------------------------------------
% Correspondence
\vspace{0.5cm}
\noindent \textbf{Correspondence to:} Areejit Samal (asamal@imsc.res.in ) 
or Anirban Chakraborti (anirban@jnu.ac.in)
%-----------------------------------------------------------------------------

%-----------------------------------------------------------------------------
% References
%merlin.mbs apsrev4-1.bst 2010-07-25 4.21a (PWD, AO, DPC) hacked
%Control: key (0)
%Control: author (8) initials jnrlst
%Control: editor formatted (1) identically to author
%Control: production of article title (-1) disabled
%Control: page (0) single
%Control: year (1) truncated
%Control: production of eprint (0) enabled
%
%-----------------------------------------------------------------------------

%-----------------------------------------------------------------------------
\newpage
%-----------------------------------------------------------------------------
% Supplementary Material
\section*{Supplementary Material}
\setcounter{table}{0}
\renewcommand{\thetable}{S\arabic{table}}%
\setcounter{figure}{0}
\renewcommand{\thefigure}{S\arabic{figure}}%
%-----------------------------------------------------------------------------

%--------------------------------------------------------------------------
% Supplementary Text
\section*{Supplementary Text}
%--------------------------------------------------------------------------

%--------------------------------------------------------------------------
% Markowitz portfolio optimization
\section*{Markowitz portfolio optimization}

We computed the risk corresponding to the portfolio comprising the market
indices using the Markowitz framework as an indicator of the market risk 
for an investor who wishes to maximize the expected returns with the 
constraint of minimum variance. That is, the scheme minimizes 
$\mathrm{w}^{\prime} \Sigma \mathrm{w}- \phi R^{\prime} \mathrm{w}$ with 
respect to the normalized weight vector $\mathrm{w}$, where $\Sigma$ is 
the covariance matrix calculated from the logarithmic returns of the market 
indices, $\phi$ is the measure of risk appetite of investor and  $R^{\prime}$ 
is the expected return of the market indices. We specify short-selling 
constraint, $\phi=0$ and $\mathrm{w}_i \geq 0$, such that we get a convex 
combination of returns of market indices for finding the minimum risk 
portfolio. These computations were performed using the in-built function 
in \texttt{MATLAB Portfolio} package 
(\url{https://in.mathworks.com/help/finance/portfolio.html}).

%--------------------------------------------------------------------------

%--------------------------------------------------------------------------
% Standard network measures
\section*{Standard network measures}

Each network investigated in this work can be represented as a weighted and
undirected graph $G(V,E)$ where $V$ is the set of vertices (or nodes) and
$E$ is the set of edges (or links) in the graph. Also, an edge in a weighted 
graph has weight assigned to it, and this weight in real networks typically 
represents the distance or strength between vertices forming the edge. 
Depending on the network measure employed for characterizing the weighted 
graph, either the strength or the distance between two vertices could be the 
appropriate natural weight to use in the associated computation (Supplementary 
Table \ref{measure_table}). Recall that while the strength represents 
similarity between two vertices, the distance reflects dissimilarity between 
them. 

Here, we have studied weighted networks constructed from cross-correlation 
among global financial market indices (see Methods section in main text). For 
these networks of global market indices, we use the absolute value of the 
correlation $C_{ij}^{\tau}(t)$, that is $|C_{ij}^{\tau}(t)|$, between two 
market indices $i$ and $j$ as the strength of the edge between vertices $i$ 
and $j$ in the threshold network for epoch ending at $t$ for computations, 
and the ultrametric distance $D_{ij}^{\tau}(t)=\sqrt{2(1-C_{ij}^{\tau}(t))}$ 
as the distance of the edge between vertices $i$ and $j$ for computations 
(Supplementary Table \ref{measure_table}). Note that the strength of an edge 
given by $|C_{ij}^{\tau}(t)|$ in the network of global financial market 
indices can take a value between 0 and 1, while the distance of an edge given 
by $D_{ij}^{\tau}(t)$ can take a value between 0 and 2. 
   
In this work, we have characterized the structure of the global market
indices network represented as a weighted and undirected graph using the
following measures.
\begin{itemize}

\item The \textit{number of edges} $m$ is given by $m = |E|$ and the number
of vertices $n$ is given by $n = |V|$, where $|\ |$ denotes cardinality of the
set.

\item The number of edges incident on a given vertex gives its degree. The
\textit{average degree} of vertices can be expressed as
$ \langle k \rangle =\frac{2 m}{n}$, where $n$ is the number of vertices and
$m$ is the number of edges in graph $G$.

\item The weighted degree (or strength) of a vertex is defined as the sum of
weights of the edges incident on the vertex \citep{Barrat2004}. Consequently,
the \textit{average weighted degree} of vertices can be defined as
$ \langle k_w \rangle=\frac{2 m_w}{n}$, where $m_w$ is the sum of weights
assigned to all edges in graph $G$. We remark that the strength is the natural
edge weight while computing the average weighted degree. In the main text, we 
also sometimes refer to average weighted degree as \textit{average strength}.

\item The \textit{edge density} is defined as the ratio of the number of edges
and the number of possible edges in graph $G$. Since a total of $\frac{n(n-1)}{2}$
edges are possible in an undirected graph $G$ ignoring self-edges, the edge
density is given by $\frac{2m}{n(n-1)}$.

\item The \textit{shortest path} between any two vertices $i$ and $j$ in a graph 
is defined as a path wherein the sum of the distance along all the edges in the 
path is the minimum among all possible paths connecting the two vertices. The 
\textit{shortest path length}, denoted by $d(i,j)$, is the sum of distances along 
edges in the shortest path between vertices $i$ and $j$ in the graph. The 
\textit{average (shortest) path length} is an average of the shortest path lengths 
between every pair of vertices in the graph, that is,
\begin{equation}
\langle L \rangle =\frac{1}{n(n-1)} \sum_{i\neq j \in V} d(i,j).
\end{equation}
We remark that distance is the natural edge weight while computing the average 
shortest path length in weighted graphs.

\item The \textit{diameter} of a graph is defined as the maximum of the shortest 
paths between all pairs of vertices, that is, $max\{ d(i,j)\ \forall i,j \in V \}$.

\item \textit{Communcation efficiency} characterizes the global information flow or 
ability to exchange information in a network \citep{Latora2001}. The communication 
efficiency $\mu_c$ is defined as
\begin{equation}
\mu_c = \frac{1}{n(n-1)} \sum_{i\neq j \in V} \frac{1}{d(i,j)}.
\end{equation}
We remark that distance is the natural edge weight while computing the communication 
efficiency in weighted graphs.

\item The \textit{clustering coefficient} of a vertex gives a measure of its tendency to
form triads with its neighbouring vertices. Onnela \cite{Onnela2005} has proposed an
approach to measure the clustering coefficient in weighted networks. For a vertex $i$ in
weighted graph $G$, clustering coefficient is defined as
\begin{equation}
C_i = \frac{2}{k_i(k_i-1)}\sum_{j,k}(a_{ij}a_{ik}a_{jk})^{1/3},
\end{equation}
where $j$ and $k$ are the neighbours of vertex $i$ and the summation runs over all such
pairs of neighbours. The quantity in the summation is the intensity of the triangle attached
to vertex $i$, and it takes the value $0$ if a triangle is not formed. The \textit{average
clustering coefficent} of a graph $G$ is the average of the clustering coefficients across
all vertices in $G$. We remark that the strength is the natural edge weight while computing
the clustering coefficient in weighted graphs.

\item  A network is said to exhibit community structure if it is possible to divide
the vertices into distinct groups of densely connected vertices. Modularity measures
edge density within a community in comparison to the edges between communities.
Modularity of a weighted graph $G$ is defined as \citep{Girvan2002,Blondel2008}
\begin{equation}
Q=\frac{1}{2m_w}\sum_{i\neq j \in V} [\, a_{ij} - \frac{s_{i}s_{j}}{2m_w}] \delta(c_i,c_j)\,
\end{equation}
where $s_i$ and $s_j$ give the sum of weights of edges attached to vertices $i$ and $j$,
respectively, $c_i$ and $c_j$ are the communities of $i$ and $j$, respectively, and $m_w$
is the sum of weights of all edges in $G$. We remark that the strength is the natural
edge weight while computing the modularity in weighted graphs.

\item Assortative mixing refers to the tendency of a vertex to attach to other
vertices with similar properties in the network. A network is said to be
assortative if high degree vertices tend to link with other high degree vertices.
The assortativity coefficient was introduced by Newman \citep{Newman2003} to
measure degree correlations between vertices in an unweighted network. It is
possible to extend this definition to weighted graphs by measuring how strongly
any two vertices with similar degree tend to link with each other \citep{Leung2007}.
The \textit{global assortativity} of a weighted graph $G$ is defined as:
\begin{equation}
r^w = \frac{ \frac{1}{m_w} \sum_e a_{ij} k_i k_j -
\left[ \frac{1}{2m_w} \sum_e a_{ij} (k_i + k_j) \right]^{2} }
{\frac{1}{2m_w} \sum_e a_{ij} (k_i^2 + k_j^2) -
\left[ \frac{1}{2m_w} \sum_e a_{ij} (k_i + k_j) \right]^{2} },
\end{equation}
where $m_w$ is the sum of weights of all edges, $k_i$ is the degree of the vertex
$i$, $a_{ij}$ is the weight of edge between vertices $i$ and $j$, and the summation
runs over all edges $e$ in weighted graph $G$. We remark that the strength is the
natural edge weight while computing the global assortativity in weighted graphs.

\item Measures for assortative mixing are limited since they quantify only the
linear dependence. \textit{Network entropy} was introduced to measure a network's
heterogeneity \citep{Sole2004}, which follows a more general information-theoretic
approach. The remaining (excess) degree of a vertex is defined as the number of
edges leaving the vertex other than the one used to reach the vertex. The probability
that a randomly chosen vertex has an excess degree $k$ is given by the remaining
degree distribution $q_k = \frac{(k+1)p_{k+1}}{<k>}$. The network entropy $H(q)$ of
a graph $G$ is then defined as
\begin{equation}
H(q) = - \sum_{k} q_k log(q_k).
\end{equation}

\item The \textit{global reaching centrality} (GRC) is a global network measure 
that aims to quantify hierarchy in complex networks \citep{Mones2012}. This 
measure can provide information on hierarchical organization in networks. GRC 
was introduced for unweighted and undirected graphs \citep{Mones2012}, and this 
measure can be extended to weighted graphs as follows. For an edge weighted 
graph $G$, we have
\begin{equation}
GRC = \frac{\sum_{i \in V} \left [ C^{max} - C(i)) \right ]}{n-1},
\end{equation}
where $C(i)$ is the \textit{local reaching centrality} (LNC) \citep{Mones2012} of 
vertex $i$, and $C^{max}$ is the maximum value of LNC across $n$ vertices in $G$. 
The LNC for a vertex $i$ in graph G is given by
\begin{equation}
C(i) = \frac{1}{n-1} \sum_{j \in V \setminus \left \{ i \right \}} \frac{1}{d(i,j)}.
\end{equation}
In the above equation, $d(i,j)$ is the previously defined shortest path length. 
The above equation is similar to the closeness centrality for weighted networks 
with disconnected components \citep{Opsahl2010}.

\item The \textit{clique number} is defined as the size of the maximal clique
appearing in graph $G$. A clique $C$ in a graph $G(V,E)$ is a subset of the
vertices, $C \subseteq V$, such that the induced subgraph is a complete graph.

\item The centrality score of a vertex quantifies the relative importance of 
that vertex in the network. Degree centrality of a vertex, which is equal to 
its degree, is the simplest measure of centrality. Hence, a vertex can be 
considered important (or central) if it has a high degree. However, a vertex 
with low degree yet with edges to other important vertices is also an important 
vertex in the network, and this property can be accounted for by using the 
concept of \textit{eigenvector centrality}. For a vertex $i$ in a weighted 
graph $G$, its eigenvector centrality $x_i$ is defined as the weighted sum of 
the centralities of its neighbours $i$ \citep{Newman2004,Negre2018}, that is
\begin{equation}
x_i = \lambda^{-1} \sum_j a_{ij} x_j.
\end{equation}
The above equation can be rewritten as an eigenvector equation 
$\bf{Ax} = \lambda \bf{x}$, where $\bf{A}$ is the adjacency matrix of graph $G$, 
$\lambda$ is the largest eigenvalue of $\bf{A}$, and $\bf{x}$ is the eigenvector 
associated with $\lambda$. Thus, $x_i$ is the $i$\textsuperscript{th} component 
of the eigenvector $\bf{x}$.

\end{itemize}

Supplementary Table \ref{measure_table} gives an exhaustive list of network 
measures investigated here. In the table, we provide information on the type 
of edge weight, that is, strength or distance, used to compute each measure 
in the network of global market indices. The above-mentioned network measures 
were computed in networks of global market indices using programs written in 
\texttt{python} employing package \texttt{NetworkX} \citep{Hagberg2008}.

%--------------------------------------------------------------------------

%--------------------------------------------------------------------------
% Edge-based curvature measures
\section*{Edge-based curvature measures}

The Ricci curvature in differential geometry is applicable to smooth manifolds 
\citep{Jost2017}. As the classical definition of Ricci curvature is not directly 
applicable to discrete objects including graphs or networks, multiple discrete 
notions of Ricci curvature have been proposed to date \citep{Samal2018}. While 
the classical definition of Ricci curvature is associated to vectors in smooth 
manifolds, in the case of discrete networks, the Ricci curvature is naturally 
associated to edges in the graph \citep{Samal2018}. Thus, the discrete Ricci 
curvatures are associated to edges rather than vertices or nodes in a graph. In 
other words, the discrete Ricci curvatures can be employed for edge-based 
analysis in contrast to commonly used measures such as degree and clustering 
coefficient which are suited for node-based analysis of networks 
\citep{Sreejith2016,Samal2018}.  

Recall that the classical notion of Ricci curvature captures two essential
geometric properties of the manifold, namely, volume growth and dispersion of 
geodesics. However, the discretizations of Ricci curvature which have been 
employed to characterize the structure of networks cannot capture the entire
spectrum of geometric properties of the classical notion \citep{Samal2018}.
Thus, different notions of discrete Ricci curvatures may capture different 
aspects of the structure of complex networks. In this section, we describe 
four notions of discrete Ricci curvature that we have used to study the 
networks of global market indices.

%--------------------------------------------------------------------------

%--------------------------------------------------------------------------
% Ollivier-Ricci curvature
\subsection*{Ollivier-Ricci curvature}

\noindent \cite{Ollivier2007,Ollivier2009} has proposed a discrete notion of Ricci 
curvature which captures the volume growth property of the classical 
definition. Olivier's proposal is based on the observation that in spaces 
of positive (negative) curvature, balls are closer (farther) to each other
on the average than their centres. The \textit{Ollivier-Ricci curvature} 
(OR) of an edge $e$ between vertices $i$ and $j$ in undirected graph $G$ 
is defined as
\begin{equation}
\mathbf{O}(e)  = 1 - \frac{W_1(m_i,m_j)}{d(i,j)} \ ,
\end{equation}
where $m_i$ and $m_j$ are discrete probability measures assigned to vertices 
$i$ and $j$, respectively, $d(i,j)$ is the distance between $i$ and $j$, 
as defined in the previous section, and $W_1$ denotes the Wasserstein distance 
\citep{Vaserstein1969}, which is the transportation distance between $m_i$ and
$m_j$, given by
\begin{equation}
W_1(m_i, m_j)=\inf_{\mu_{i,j}\in \prod(m_i, m_j)}\sum_{(i',j')\in V\times V}
d(i', j')\mu_{i,j}(i', j'),
\end{equation}
where $\prod(m_i, m_j)$ is the set of probability measures $\mu_{i,j}$ that
satisfy
\begin{equation}
\sum_{j'\in V}\mu_{i,j}(i', j')=m_i(i'), \,\,\sum_{i'\in V}\mu_{i,j}(i', j') = m_j(j').
\end{equation}
The probability distribution $m_i$ is taken to be uniform over the the 
neighbouring vertices of $i$ \citep{Lin2011}. We have computed the average 
OR curvature of edges in networks of global market indices in this work. 
While computing the OR curvature in networks of global market indices, 
the weight of each edge is taken to be the distance. Given the OR curvature 
of edges in the graph, it is straightforward to define the OR curvature of 
a vertex $v$ as
\begin{equation}
\label{FormanRicciNode}
\mathbf{O}(v) = \sum_{e\ \sim\ v} \mathbf{O}(e)\,
\end{equation}
where $e \sim v$ is the set of edges $e$ incident on vertex $v$. The above 
definition of OR curvature of a vertex is analogous to scalar curvature in
Riemannian geometry \citep{Samal2018}.

%--------------------------------------------------------------------------

%--------------------------------------------------------------------------
% Forman-Ricci curvature
\subsection*{Forman-Ricci curvature}

\noindent Forman's discretization \citep{Forman2003} captures the geodesic dispersal 
property of the classical notion of Ricci curvature \citep{Sreejith2016}. It 
is based on the relation between the Riemann-Laplace operator and Ricci 
curvature. Recently, \textit{Forman-Ricci curvature} (FR) was adapted for the 
analysis of  unweighted and weighted networks \citep{Sreejith2016,Sreejith2017}. 
Intuitively, FR curvature quantifies the information spread at the ends of 
an edge in the network. High negative FR value for an edge indicates more 
spread of information at its ends. For an edge $e$ between vertices $i$ and $j$ 
in an undirected graph $G$, FR is defined as
\begin{equation}
\label{FormanRicciEdge}
\mathbf{F}(e) = w_e \left( \frac{w_{i}}{w_e} +  \frac{w_{j}}{w_e}
- \sum_{e_{i}\ \sim\ e,\ e_{j}\ \sim\ e}
\left[\frac{w_{i}}{\sqrt{w_e w_{e_{j} }}}
+ \frac{w_{j}}{\sqrt{w_e w_{e_{j} }}} \right] \right)\,
\end{equation}
where $w_e$ denotes the weight of the edge $e$, $w_{i}$ and $w_{j}$ denote 
the weights associated with the vertices $i$ and $j$, respectively, $e_{i} 
\sim e$ and $e_{j} \sim e$ denote the set of edges incident on vertices $i$ 
and $j$, respectively, after excluding the edge $e$. We have computed the 
average FR curvature of edges in networks of global market indices in this
work. While computing the FR curvature in networks of global market indices, 
the weight of each edge is taken to be the distance while the weight of each 
vertex is taken to be $1$.

Given the FR curvature of edges in the graph, it is straightforward to define 
the FR curvature of a vertex $v$ as
\begin{equation}
\label{FormanRicciNode}
\mathbf{F}(v) = \sum_{e\ \sim\ v} \mathbf{F}(e)\,
\end{equation}
where $e \sim v$ is the set of edges $e$ incident on vertex $v$. The above 
definition of FR curvature of a vertex is analogous to scalar curvature in
Riemannian geometry \citep{Sreejith2017}.
 
%--------------------------------------------------------------------------

%--------------------------------------------------------------------------
% Menger-Ricci curvature
\subsection*{Menger-Ricci curvature}

\noindent Menger defined the curvature of a metric triangle $T$ \citep{Menger1930} 
formed by three points in space as the reciprocal $\frac{1}{R(T)}$ of the radius 
$R(T)$ of the circumcircle of that triangle. Given a triangle $T = T(a,b,c)$ 
with sides $a,\ b,\ c$ in a metric space $(M,d)$, the Menger curvature of 
$T$ is given by
\begin{equation}
K_M(T) = \frac{\sqrt{p(p - a)(p - b)(p - c)}}{a \cdot b \cdot c}\,
\end{equation}
where $p = (a + b + c)/2$. It is possible to extend the above definition 
to unweighted and undirected networks \citep{Saucan2019b,Saucan2020}, where 
one considers combinatorial triangles with length of each side equal to $1$,
and this gives $K_M(T)=\sqrt{3}/2$. Then the \textit{Menger-Ricci curvature} 
(MR) of an edge $e$ in the graph $G$ can be defined as
\begin{equation}
\label{MengerRicciEdge}
\kappa_M(e) = \sum_{T_e \sim e}\kappa_{M}(T_e)\,,
\end{equation}
where $T_e \sim e$ denote the triangles adjacent to the edge $e$. An edge 
will have high positive value of MR curvature if it is part of many triangles 
in the network. In this work, we have computed the average MR curvature of 
edges in networks of global market indices.

%--------------------------------------------------------------------------

%--------------------------------------------------------------------------
% Haantjes-Ricci curvature
\subsection*{Haantjes-Ricci curvature}

\noindent Haantjes \citep{Haantjes1947} defined the curvature of a metric curve as the 
ratio of the length of the arc of the curve and that of the chord it subtends. 
More precisely, given three points $p$, $q$ and $r$ on a curve in a metric 
space such that $p$ lies between $q$ and $r$, the Haantjes curvature at the 
point $p$ is defined as
\begin{equation}
\kappa_{H}^2(p) = 24\lim_{q,r \rightarrow p}\frac{l(\widehat{qr})-d(q,r)}
{\big(d(q,r)\big)^3}\,,
\end{equation}
where $l(\widehat{qr})$ denotes the length of the arc $\widehat{qr}$. The above 
definition can be extended to networks by replacing the arc $\widehat{qr}$ with 
a path between the two vertices and the subtending chord by the edge between the 
two vertices \citep{Saucan2019b,Saucan2020}. Given a simple path $\pi = i,\ldots,j$ 
between the two vertices $i$ and $j$ connected by an edge $e$ in the unweighted 
graph $G$, the Haantjes curvature of the path takes the value
\begin{equation}
\kappa_H(\pi) = \sqrt{n-1} \ ,
\end{equation}
where $n$ is the number of edges appearing in the path $\pi$. Then the 
\textit{Haantjes-Ricci curvature} (HR) of the edge $e$ can be defined as 
\citep{Saucan2019b,Saucan2020}
\begin{equation}
\kappa_H(e) = \sum_{\pi \sim e}\kappa_{H}(\pi)\,,
\end{equation}
where the summation runs over all the paths between vertices $i$ and $j$.
In this work, we have computed the average HR curvature of edges in networks 
of global market indices by ignoring edge weights. Further, computational 
constraints permitted only consideration of paths $\pi$ of length $\leq 5$
between two vertices at the ends of any edge while computing the HR curvature 
in networks of global market indices.

%--------------------------------------------------------------------------

%--------------------------------------------------------------------------
% Planar maximally filtered graph (PMFG) construction and characteristics
\section*{Planar maximally filtered graph (PMFG) construction and 
characteristics}

Here, we describe an alternate network construction framework, namely, the 
planar maximally filtered graph (PMFG) \citep{Tumminello2005}, which has been 
widely-used to study the relationship between global financial market indices.
Briefly, the PMFG $\mathbf{P}^{\tau}(t)$ of market indices can be constructed 
for the time-series of cross-correlation matrices $\mathbf{C}^{\tau}(t)$ of 
window size $\tau=80$ days and an overlapping shift of $\Delta\tau=20$ days 
over the 14-year period as follows (see Methods section in main text for the 
computation of cross-correlation matrices $\mathbf{C}^{\tau}(t)$ starting from 
the logarithmic returns of daily closing prices of 69 global market indices). 
Firstly, a sorted list of edges is created based on the decreasing order of 
correlation in the matrix $\mathbf{C}^{\tau}(t)$. Next, each edge in the 
sorted list is considered for inclusion in the PMFG based on the decreasing 
order of correlation. An edge between vertices $i$ and $j$ is added to PMFG, 
if and only if the resulting graph can be embedded on a sphere, i.e., it is a 
planar graph. Following this scheme, the final network obtained is a PMFG with 
$3(N-1)$ edges where $N$ is the number of vertices in the graph. Note that the 
minimum spanning tree (MST) in contrast to PMFG contains $N-1$ edges, and it 
has no loops or cliques. Importantly, in addition to the hierarchical organization 
of the MST, the PMFG also includes information on the loops and cliques involving 
upto four vertices present in the correlation matrix. We remark that PMFG is a 
special case of a more general method of network construction \citep{Tumminello2005}, 
wherein the edges are added under the topological constraint that the resulting 
graph can be embedded on a surface of genus $g=k$ where $k$ is a postive integer.
In case of PMFG, the graph is embedded on a surface of genus $g=0$ which is a 
sphere.  

In Supplementary Figures S10-S13, we show the temporal evolution and correlation 
between generic market indicators and network measures in PMFG $\mathbf{P}^{\tau}(t)$ 
of market indices constructed from cross-correlation matrices $\mathbf{C}^{\tau}(t)$ 
as described above. 

%--------------------------------------------------------------------------

%--------------------------------------------------------------------------
% Supplementary Figures
%--------------------------------------------------------------------------

%--------------------------------------------------------------------------
% Figure S1
\begin{figure}[ht]
\label{FigS1}
\begin{center}
\includegraphics[width = 0.79\linewidth]{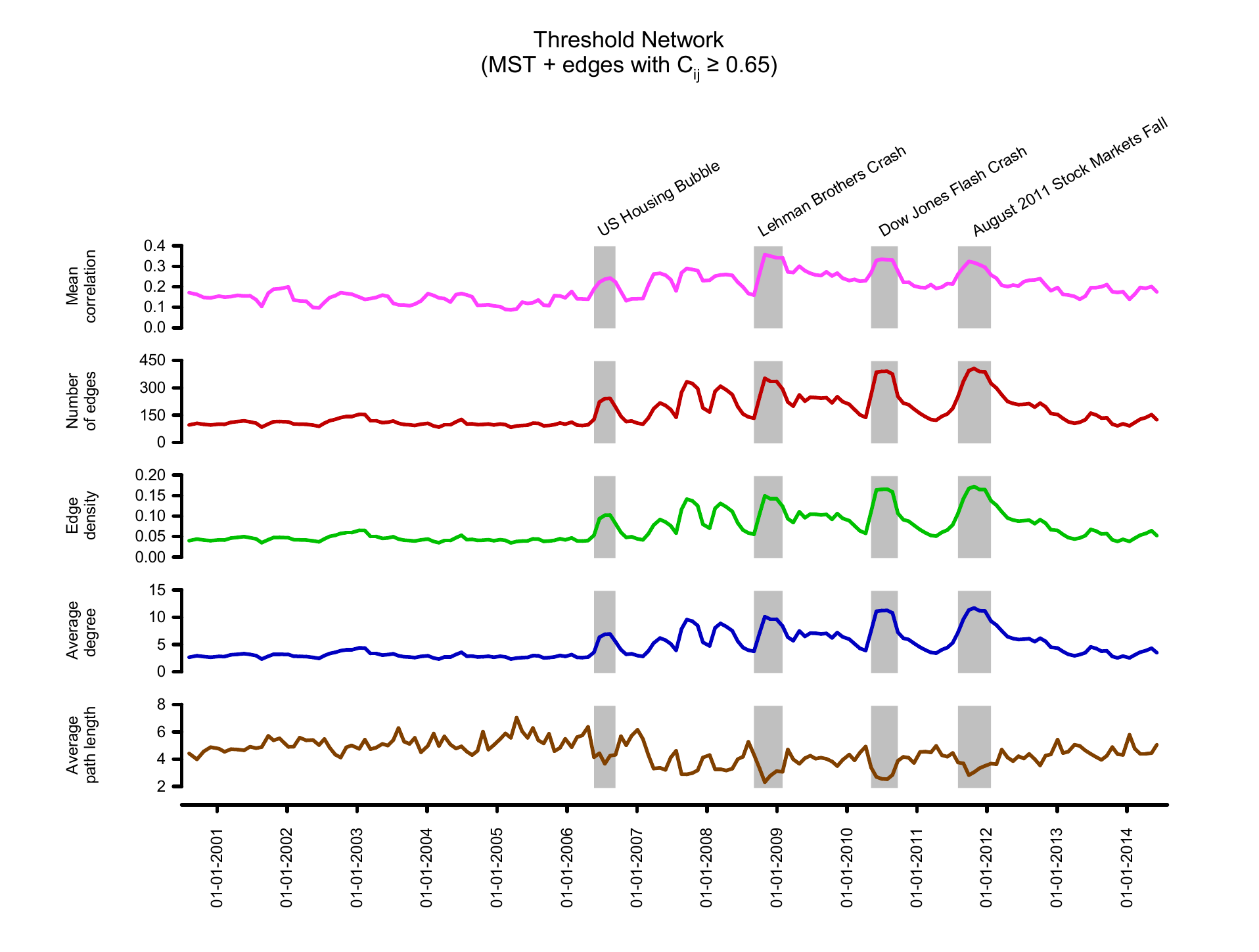}
\end{center}
\caption{Evolution of generic indicators and network characteristics for the 
global market indices networks $\mathbf{S}^{\tau}(t)$, constructed from the 
correlation matrices $\mathbf{C}^{\tau}(t)$ of window size $\tau=80$ days and 
an overlapping shift of $\Delta\tau=20$ days over a period of 14 years 
(2000-2014). The threshold networks $\mathbf{S}^{\tau}(t)$ were constructed 
by adding edges with correlation $C_{ij}^{\tau}(t) \ge 0.65$ to the minimum spanning tree 
(MST). From top to bottom, we compare the plot of mean correlation among market 
indices, number of edges, edge density, average degree and average path length. 
The four shaded regions correspond to the epochs around the four important 
market events, namely, US housing bubble, Lehman brothers crash, Dow Jones 
flash crash, and August 2011 stock markets fall.}
\end{figure}
%--------------------------------------------------------------------------

%--------------------------------------------------------------------------
% Figure S2
\begin{figure}[ht]
\label{FigS2}
\begin{center}
\includegraphics[width = 0.79\linewidth]{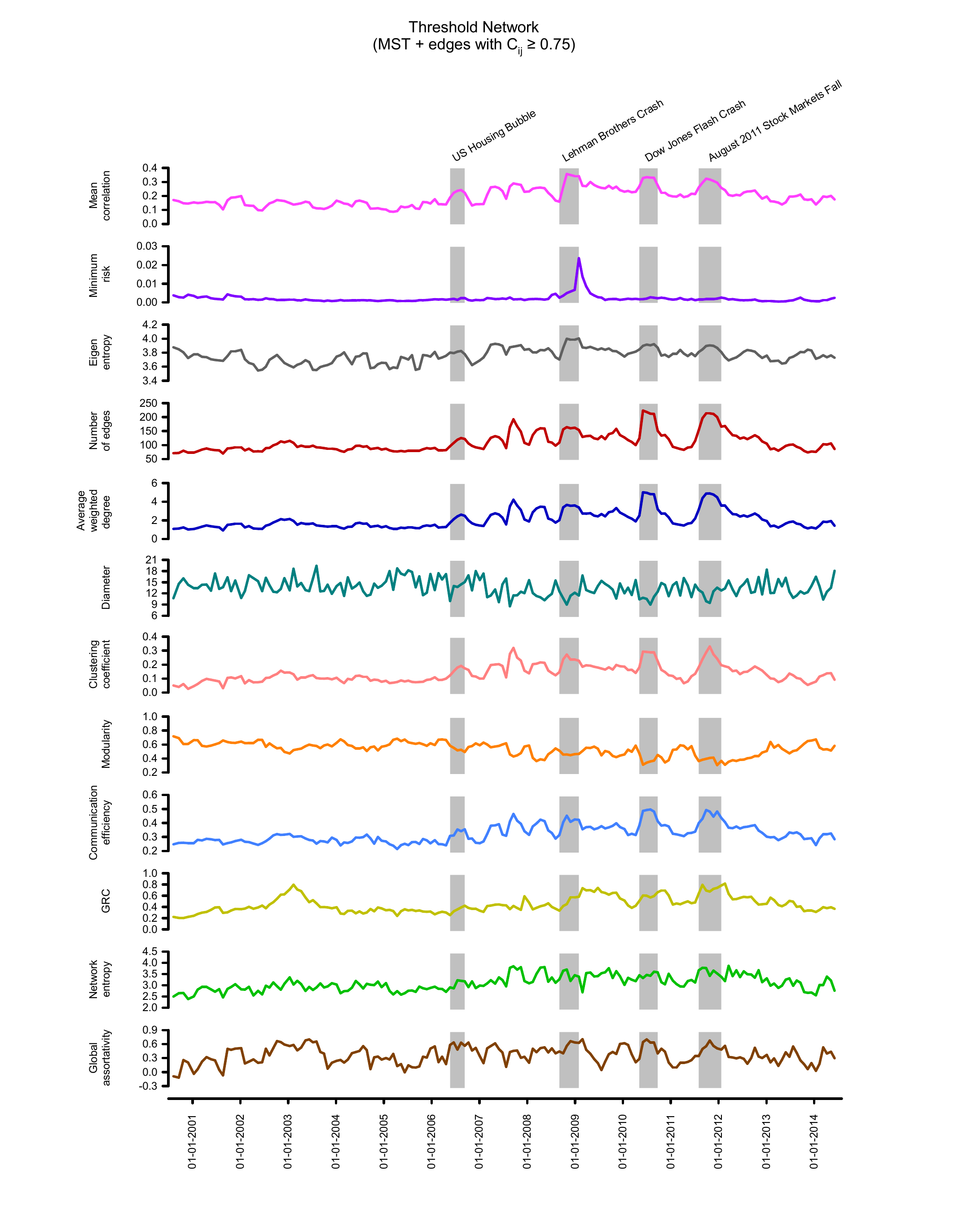}
\end{center}
\caption{Evolution of generic indicators and network characteristics for the 
global market indices networks $\mathbf{S}^{\tau}(t)$, constructed from the 
correlation matrices $\mathbf{C}^{\tau}(t)$ of window size $\tau=80$ days and 
an overlapping shift of $\Delta\tau=20$ days over a period of 14 years 
(2000-2014). The threshold networks $\mathbf{S}^{\tau}(t)$ were constructed 
by adding edges with correlation $C_{ij}^{\tau}(t) \ge 0.75$ to the minimum spanning tree 
(MST). From top to bottom, we compare the plot of mean correlation among market 
indices, minimum risk corresponding to the Markowitz portfolio optimization, 
eigen-entropy, number of edges, average weighted degree, diameter, clustering 
coefficient, modularity, communication efficiency, global reaching centrality 
(GRC), network entropy and global assortativity. The four shaded regions 
correspond to the epochs around the four important market events, namely,  
US housing bubble, Lehman brothers crash, Dow Jones flash crash, and August 
2011 stock markets fall.}
\end{figure}
%--------------------------------------------------------------------------

%--------------------------------------------------------------------------
% Figure S3
\begin{figure}[ht]
\label{FigS3}
\begin{center}
\includegraphics[width = 0.79\linewidth]{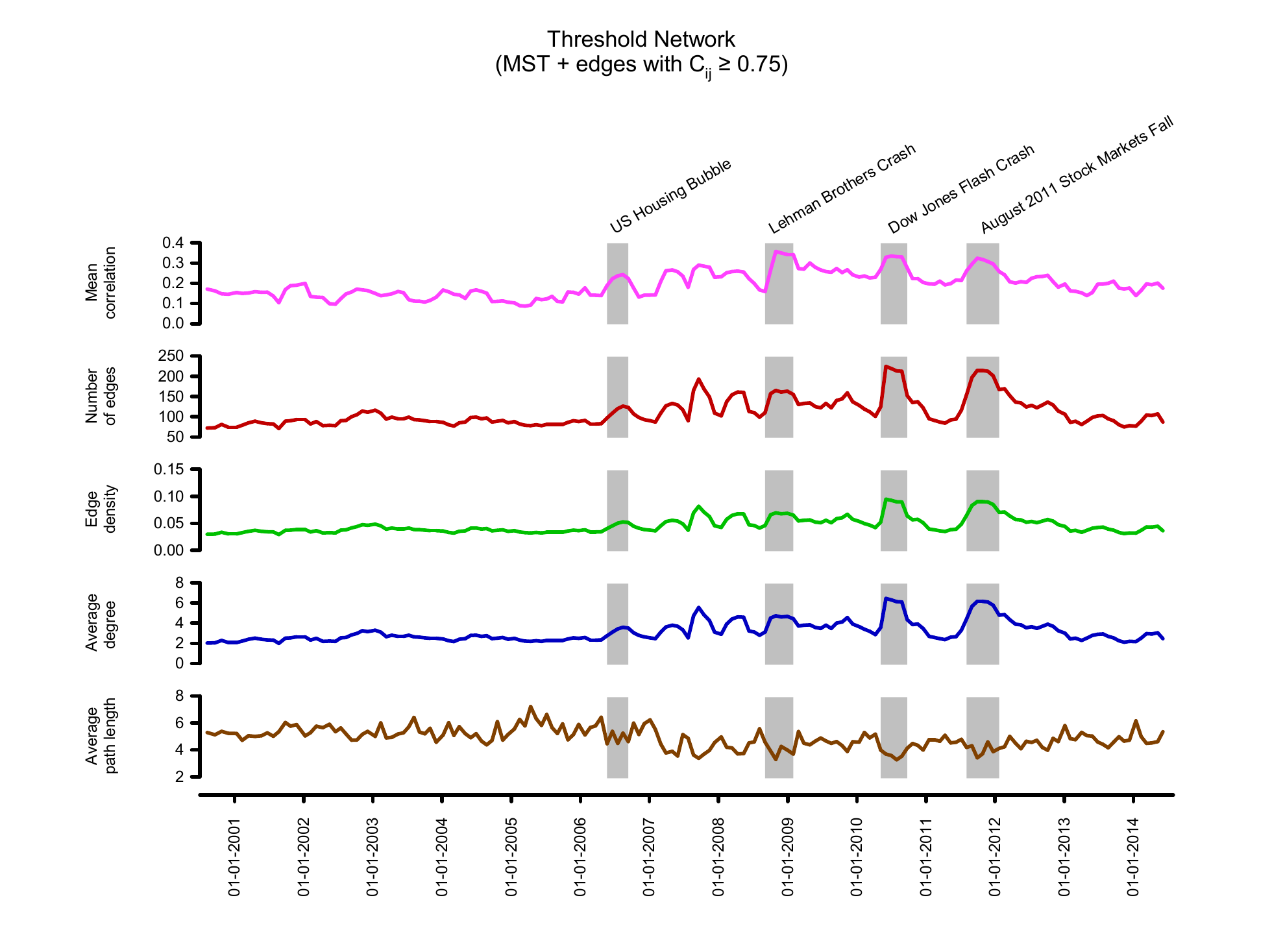}
\end{center}
\caption{Evolution of generic indicators and network characteristics for the 
global market indices networks $\mathbf{S}^{\tau}(t)$, constructed from the 
correlation matrices $\mathbf{C}^{\tau}(t)$ of window size $\tau=80$ days and 
an overlapping shift of $\Delta\tau=20$ days over a period of 14 years 
(2000-2014). The threshold networks $\mathbf{S}^{\tau}(t)$ were constructed 
by adding edges with correlation $C_{ij}^{\tau}(t) \ge 0.75$ to the minimum spanning tree 
(MST). From top to bottom, we compare the plot of mean correlation among market 
indices, number of edges, edge density, average degree and average path length. 
The four shaded regions correspond to the epochs around the four important 
market events, namely, US housing bubble, Lehman brothers crash, Dow Jones 
flash crash, and August 2011 stock markets fall.}
\end{figure}
%--------------------------------------------------------------------------

%--------------------------------------------------------------------------
% Figure S4
\begin{figure}[ht]
\label{FigS4}
\begin{center}
\includegraphics[width = 0.79\linewidth]{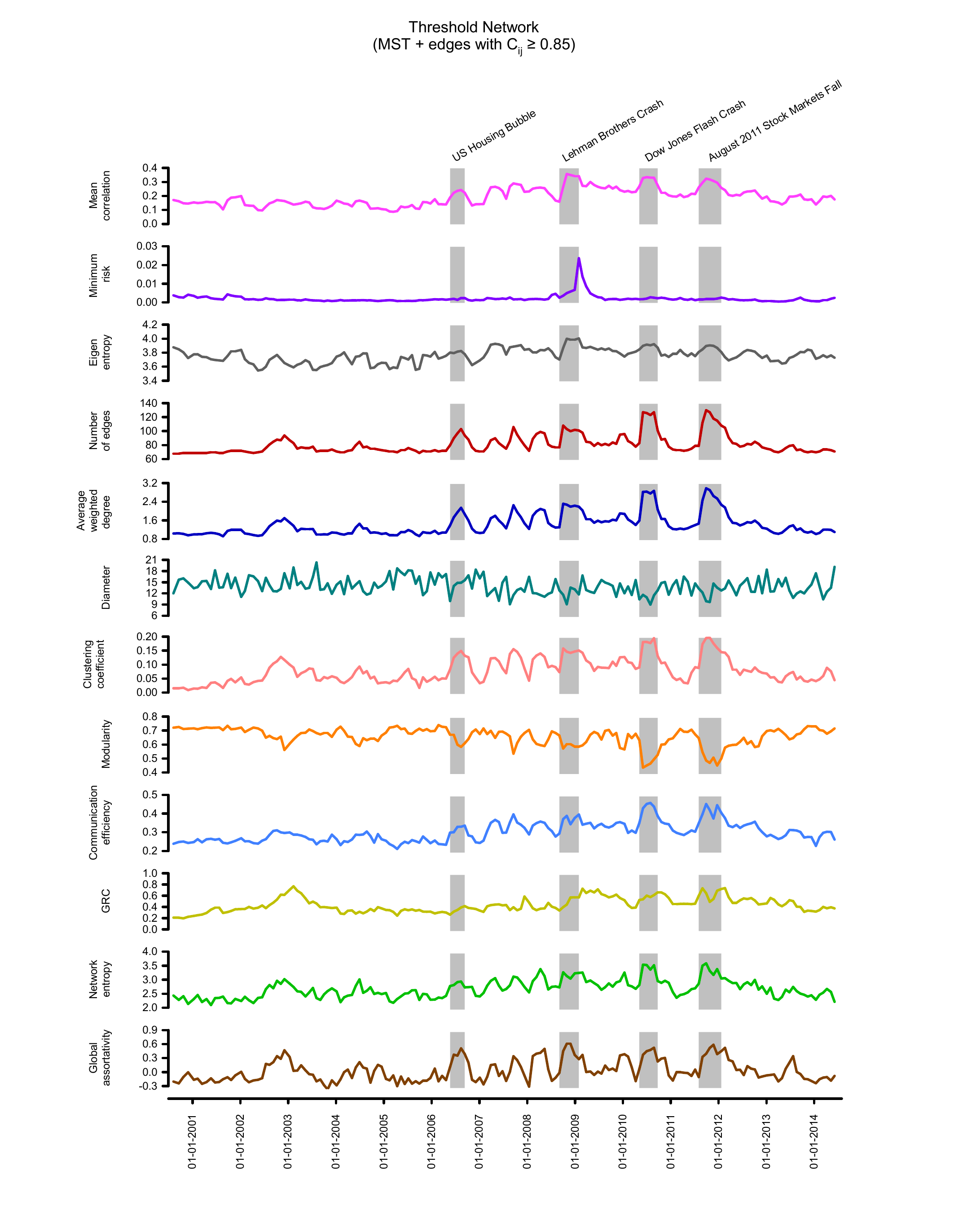}
\end{center}
\caption{Evolution of generic indicators and network characteristics for the 
global market indices networks $\mathbf{S}^{\tau}(t)$, constructed from the 
correlation matrices $\mathbf{C}^{\tau}(t)$ of window size $\tau=80$ days and 
an overlapping shift of $\Delta\tau=20$ days over a period of 14 years 
(2000-2014). The threshold networks $\mathbf{S}^{\tau}(t)$ were constructed 
by adding edges with correlation $C_{ij}^{\tau}(t) \ge 0.85$ to the minimum spanning tree 
(MST). From top to bottom, we compare the plot of mean correlation among market 
indices, minimum risk corresponding to the Markowitz portfolio optimization, 
eigen-entropy, number of edges, average weighted degree, diameter, clustering 
coefficient, modularity, communication efficiency, global reaching centrality 
(GRC), network entropy and global assortativity. The four shaded regions 
correspond to the epochs around the four important market events, namely,  
US housing bubble, Lehman brothers crash, Dow Jones flash crash, and August 
2011 stock markets fall.}
\end{figure}
%--------------------------------------------------------------------------

%--------------------------------------------------------------------------
% Figure S5
\begin{figure}[ht]
\label{FigS5}
\begin{center}
\includegraphics[width = 0.79\linewidth]{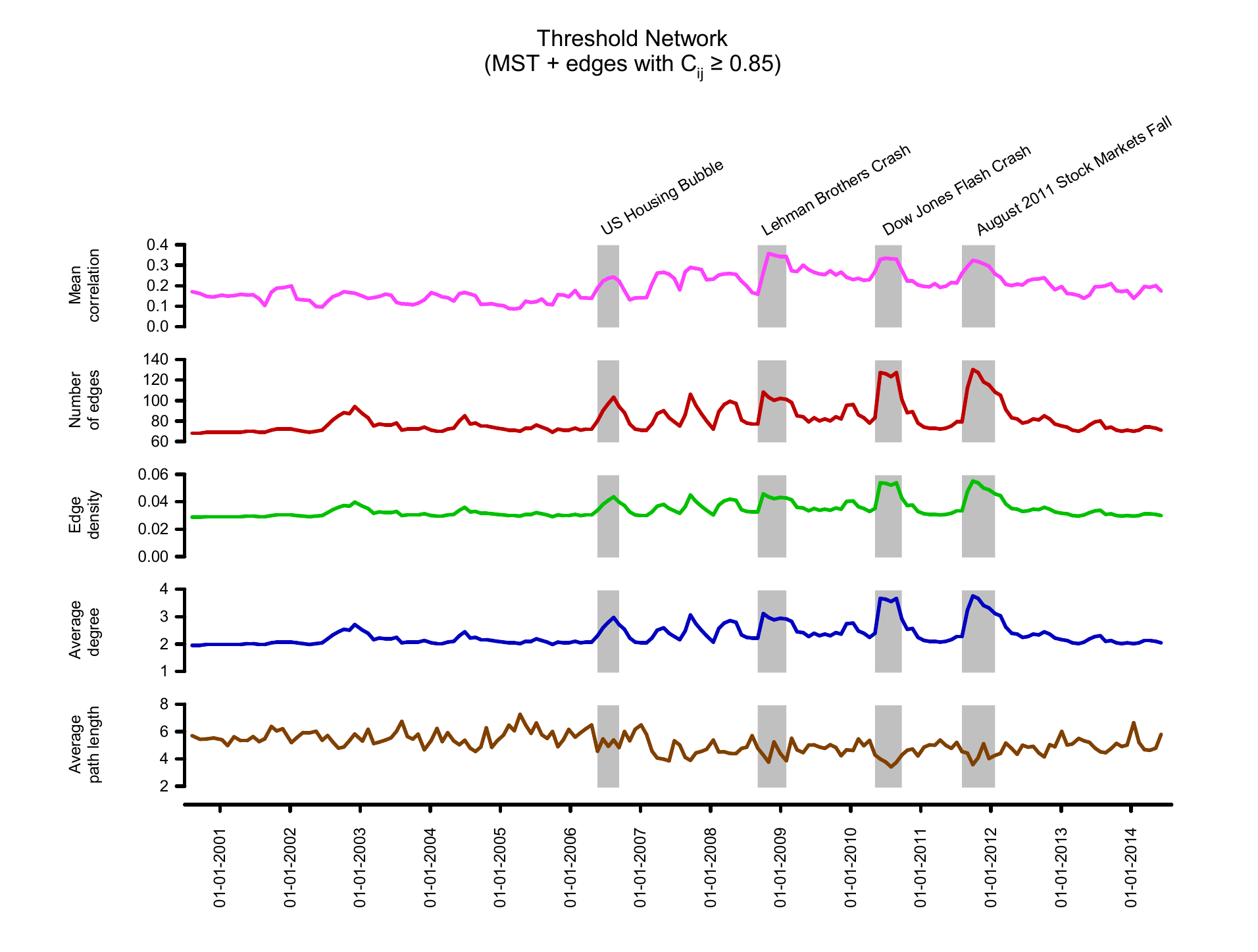}
\end{center}
\caption{Evolution of generic indicators and network characteristics for the 
global market indices networks $\mathbf{S}^{\tau}(t)$, constructed from the 
correlation matrices $\mathbf{C}^{\tau}(t)$ of window size $\tau=80$ days and 
an overlapping shift of $\Delta\tau=20$ days over a period of 14 years 
(2000-2014). The threshold networks $\mathbf{S}^{\tau}(t)$ were constructed 
by adding edges with correlation $C_{ij}^{\tau}(t) \ge 0.85$ to the minimum spanning tree 
(MST). From top to bottom, we compare the plot of mean correlation among market 
indices, number of edges, edge density, average degree and average path length. 
The four shaded regions correspond to the epochs around the four important 
market events, namely, US housing bubble, Lehman brothers crash, Dow Jones 
flash crash, and August 2011 stock markets fall.}
\end{figure}
%--------------------------------------------------------------------------

%--------------------------------------------------------------------------
% Figure S6
\begin{figure}[ht]
\label{FigS6}
\begin{center}
\includegraphics[width = 0.72\linewidth]{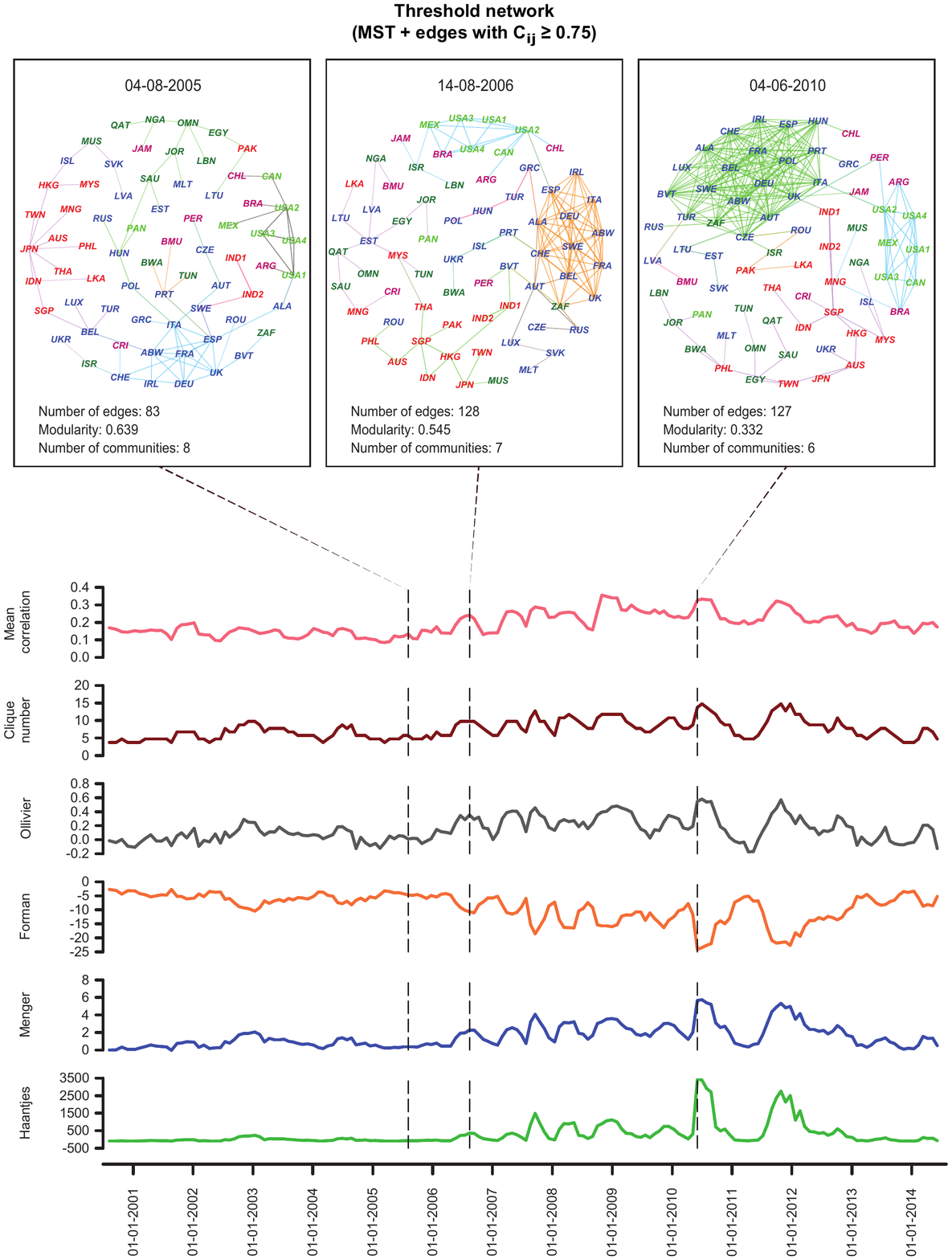}
\end{center}
\caption{Evolution of network characteristics and visualization of the 
threshold networks $\mathbf{S}^{\tau}(t)$ of market indices with window 
size $\tau=80$ days and an overlapping shift of $\Delta\tau=20$ days, 
constructed by adding edges with correlation $C_{ij}^{\tau}(t) \ge 0.75$ to the MST. 
(Lower panel) Comparison of the plots of mean correlation among market 
indices, clique number, average of Ollivier-Ricci (OR), Forman-Ricci (FR), 
Menger-Ricci (MR), and Haantjes-Ricci (HR) curvature of edges in threshold 
networks over the 14-year period. (Upper panel) Visualization of the 
threshold networks at three distinct epochs of $\tau=80$ days ending on 
trading days $t$ equal to 04-08-2005 (normal), 14-08-2006 (US housing bubble) and 
04-06-2010 (Dow Jones flash crash). Threshold networks show higher number 
of edges and lower number of communities during crisis. Correspondingly, 
there is an increase in mean correlation, clique number, average OR, MR 
and HR curvature, and decrease in average FR curvature of threshold networks 
during financial crisis. Node colours and labels are based on geographical 
region and country, respectively, of the indices and edge colours are based 
on the community determined by Louvain method. The four USA market indices, 
NASDAQ, NYSE, RUSSELL1000 and SPX, are labelled as USA1, USA2, USA3 and USA4, 
respectively, while the two Indian indices, NIFTY and SENSEX30, are 
labelled as IND1 and IND2, respectively.}
\end{figure}
%--------------------------------------------------------------------------

%--------------------------------------------------------------------------
% Figure S7
\begin{figure}[ht]
\label{FigS7}
\begin{center}
\includegraphics[width = 0.72\linewidth]{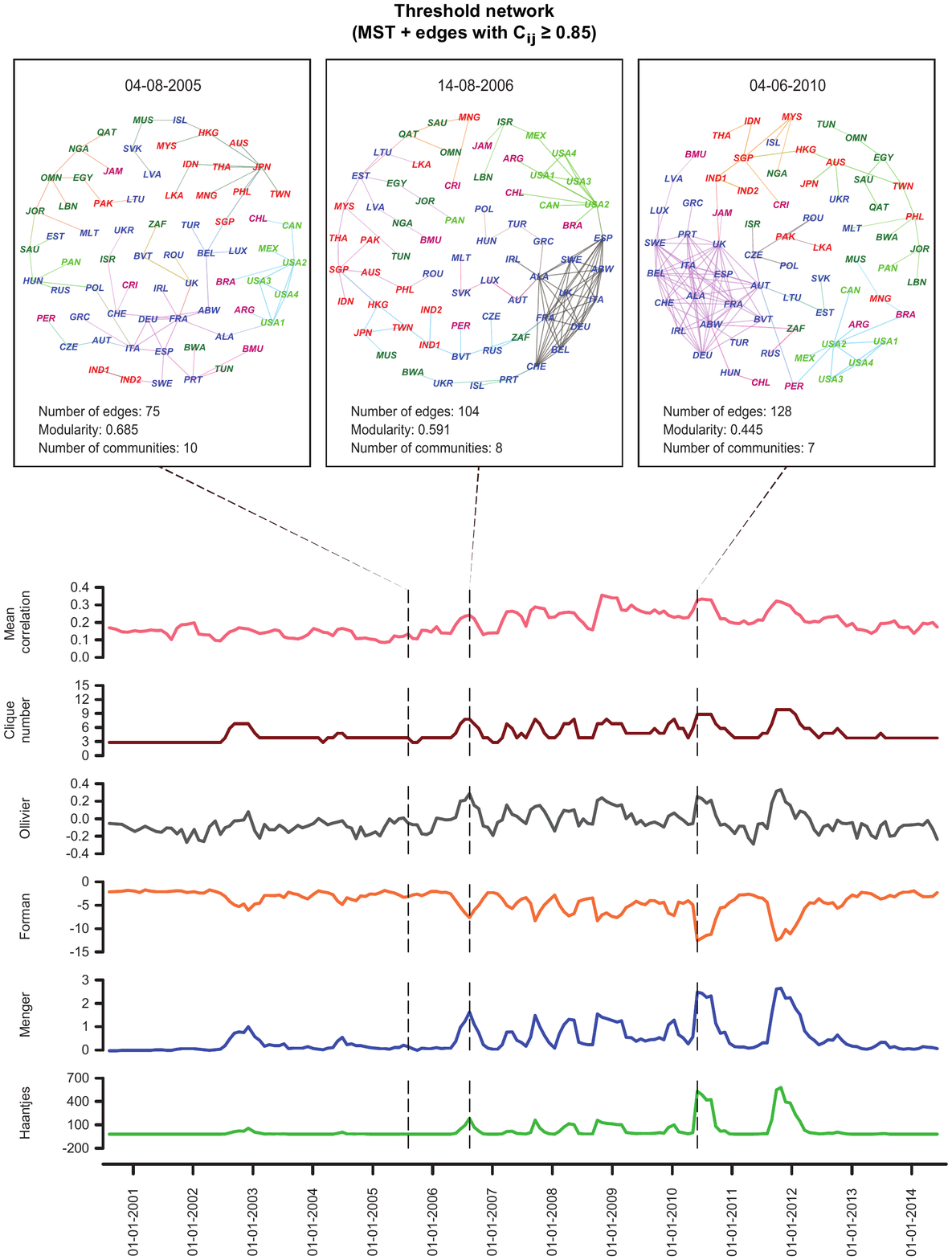}
\end{center}
\caption{Evolution of network characteristics and visualization of the 
threshold networks $\mathbf{S}^{\tau}(t)$ of market indices with window 
size $\tau=80$ days and an overlapping shift of $\Delta\tau=20$ days, 
constructed by adding edges with correlation $C_{ij}^{\tau}(t) \ge 0.85$ to the MST. 
(Lower panel) Comparison of the plots of mean correlation among market 
indices, clique number, average of Ollivier-Ricci (OR), Forman-Ricci (FR), 
Menger-Ricci (MR), and Haantjes-Ricci (HR) curvature of edges in threshold 
networks over the 14-year period. (Upper panel) Visualization of the 
threshold networks at three distinct epochs of $\tau=80$ days ending on 
trading days $t$ equal to 04-08-2005 (normal), 14-08-2006 (US housing bubble) and 
04-06-2010 (Dow Jones flash crash). Threshold networks show higher number 
of edges and lower number of communities during crisis. Correspondingly, 
there is an increase in mean correlation, clique number, average OR, MR 
and HR curvature, and decrease in average FR curvature of threshold networks 
during financial crisis. Node colours and labels are based on geographical 
region and country, respectively, of the indices and edge colours are based 
on the community determined by Louvain method. The four USA market indices, 
NASDAQ, NYSE, RUSSELL1000 and SPX, are labelled as USA1, USA2, USA3 and USA4, 
respectively, while the two Indian indices, NIFTY and SENSEX30, are 
labelled as IND1 and IND2, respectively.}
\end{figure}
%--------------------------------------------------------------------------

%--------------------------------------------------------------------------
% Figure S8
\begin{figure}[ht]
\label{FigS8}
\begin{center}
\includegraphics[width = 0.7\linewidth]{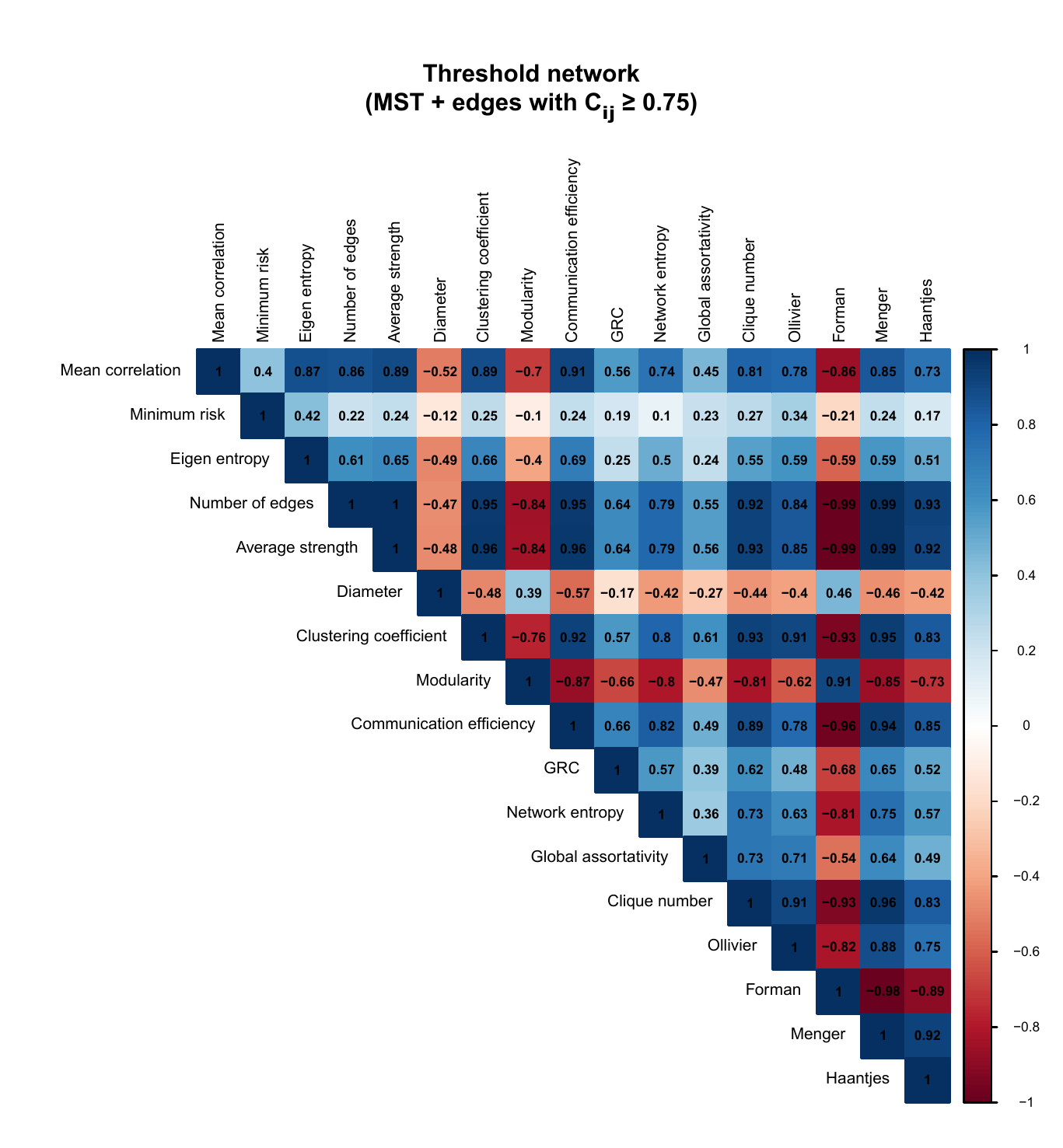}
\end{center}
\caption{Correlation between generic indicators and network characteristics 
of the global market indices networks $\mathbf{S}^{\tau}(t)$, constructed 
from the correlation matrices $\mathbf{C}^{\tau}(t)$ of window size $\tau=80$ 
days and an overlapping shift of $\Delta\tau=20$ days over a period of 14 
years (2000-2014). The threshold networks $\mathbf{S}^{\tau}(t)$ were 
constructed by adding edges with correlation $C_{ij}^{\tau}(t) \ge 0.75$ to 
the minimum spanning tree (MST).}
\end{figure}
%--------------------------------------------------------------------------

%--------------------------------------------------------------------------
% Figure S9
\begin{figure}[ht]
\label{FigS9}
\begin{center}
\includegraphics[width = 0.7\linewidth]{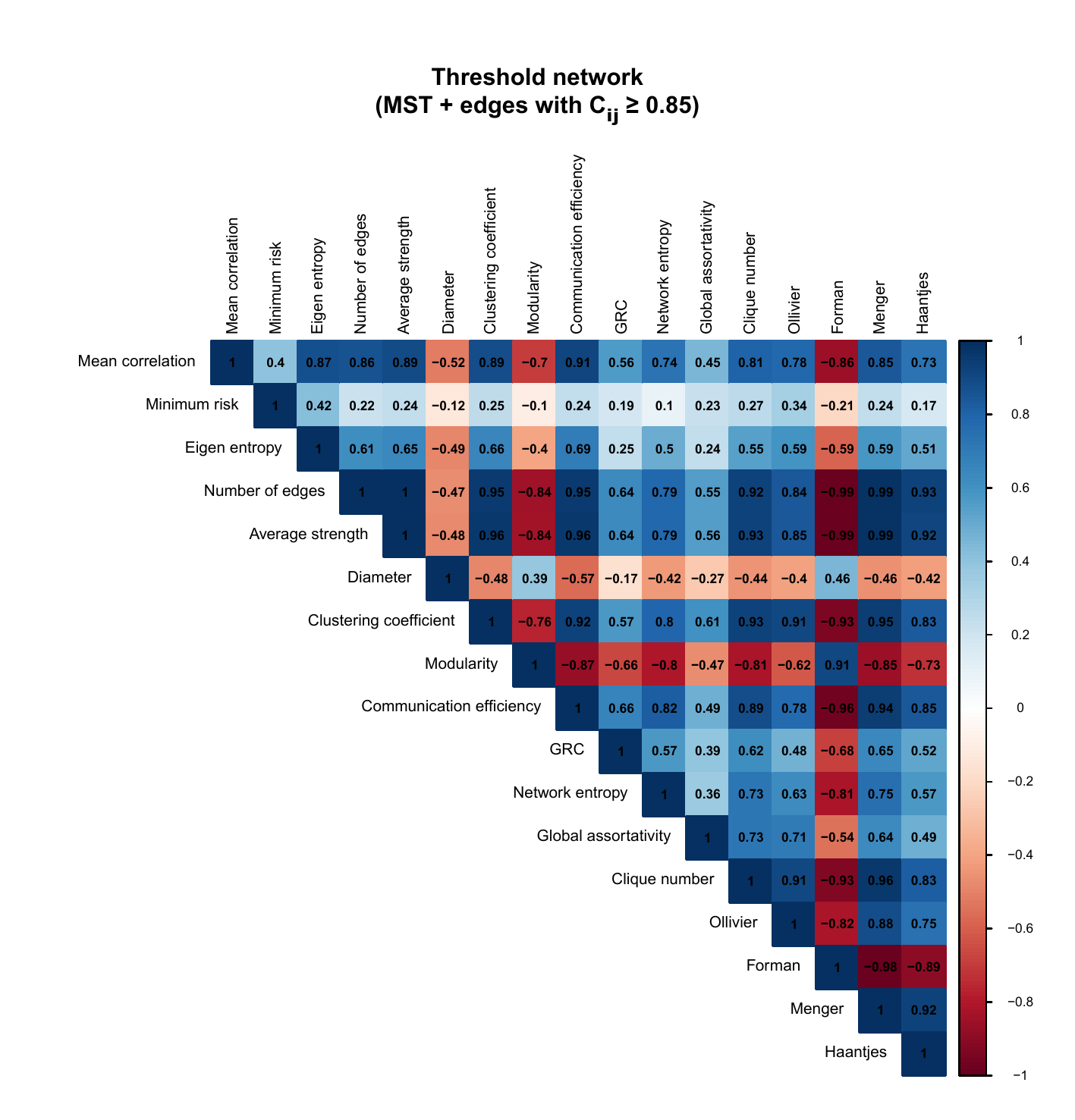}
\end{center}
\caption{Correlation between generic indicators and network characteristics 
of the global market indices networks $\mathbf{S}^{\tau}(t)$, constructed 
from the correlation matrices $\mathbf{C}^{\tau}(t)$ of window size $\tau=80$ 
days and an overlapping shift of $\Delta\tau=20$ days over a period of 14 
years (2000-2014). The threshold networks $\mathbf{S}^{\tau}(t)$ were 
constructed by adding edges with correlation $C_{ij}^{\tau}(t) \ge 0.85$ to 
the minimum spanning tree (MST).}
\end{figure}
%--------------------------------------------------------------------------

%--------------------------------------------------------------------------
% Figure S10
\begin{figure}[ht]
\label{FigS10}
\begin{center}
\includegraphics[width = 0.79\linewidth]{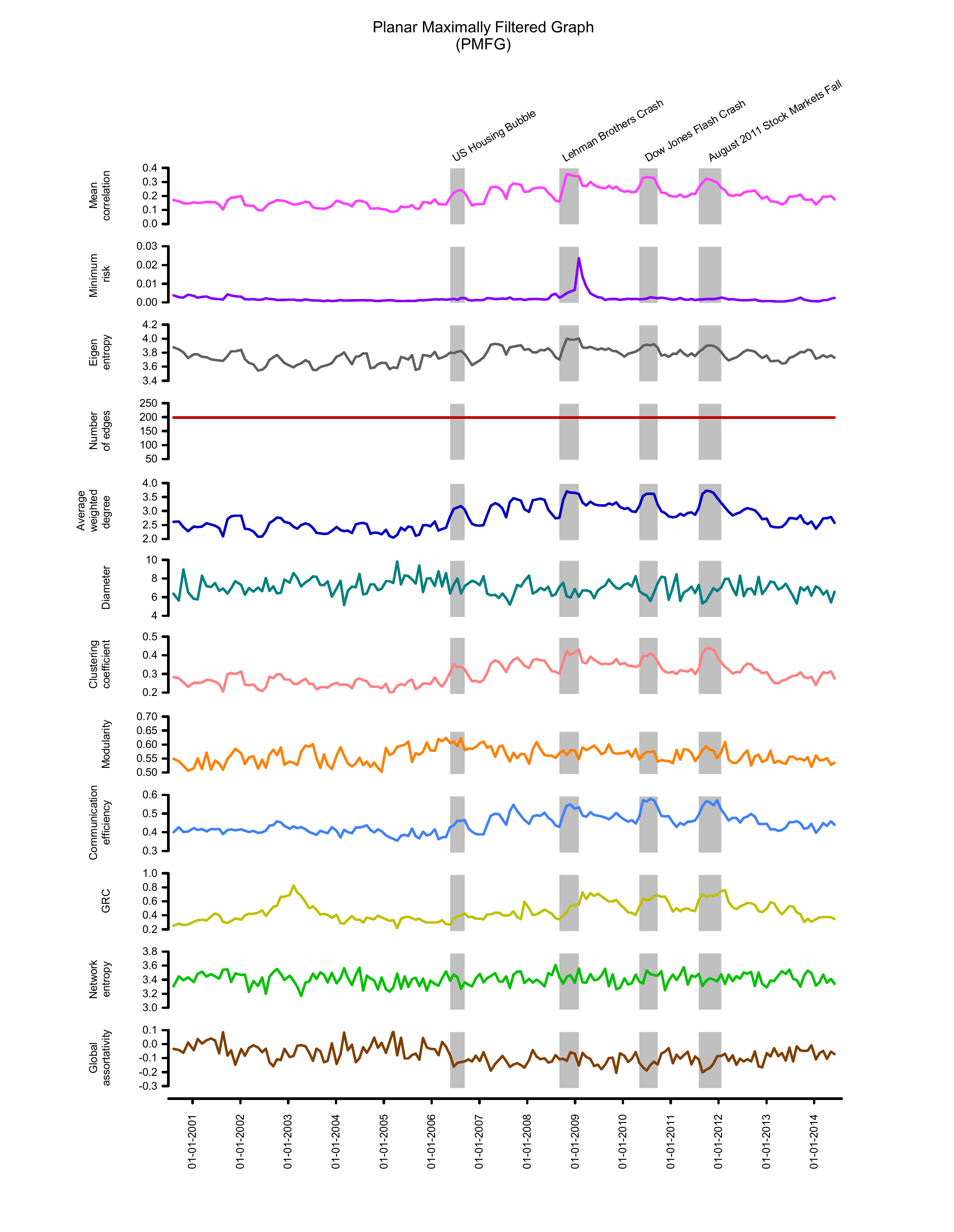}
\end{center}
\caption{Evolution of generic indicators and network characteristics for PMFG 
$\mathbf{P}^{\tau}(t)$ of global market indices, constructed from the 
correlation matrices $\mathbf{C}^{\tau}(t)$ of window size $\tau=80$ days and 
an overlapping shift of $\Delta\tau=20$ days over a period of 14 years 
(2000-2014). From top to bottom, we compare the plot of mean correlation among 
market indices, minimum risk corresponding to the Markowitz portfolio optimization, 
eigen-entropy, number of edges, average weighted degree, diameter, clustering 
coefficient, modularity, communication efficiency, global reaching centrality 
(GRC), network entropy and global assortativity. The four shaded regions 
correspond to the epochs around the four important market events, namely,  
US housing bubble, Lehman brothers crash, Dow Jones flash crash, and August 
2011 stock markets fall.}
\end{figure}
%--------------------------------------------------------------------------

%--------------------------------------------------------------------------
% Figure S11
\begin{figure}[ht]
\label{FigS11}
\begin{center}
\includegraphics[width = 0.79\linewidth]{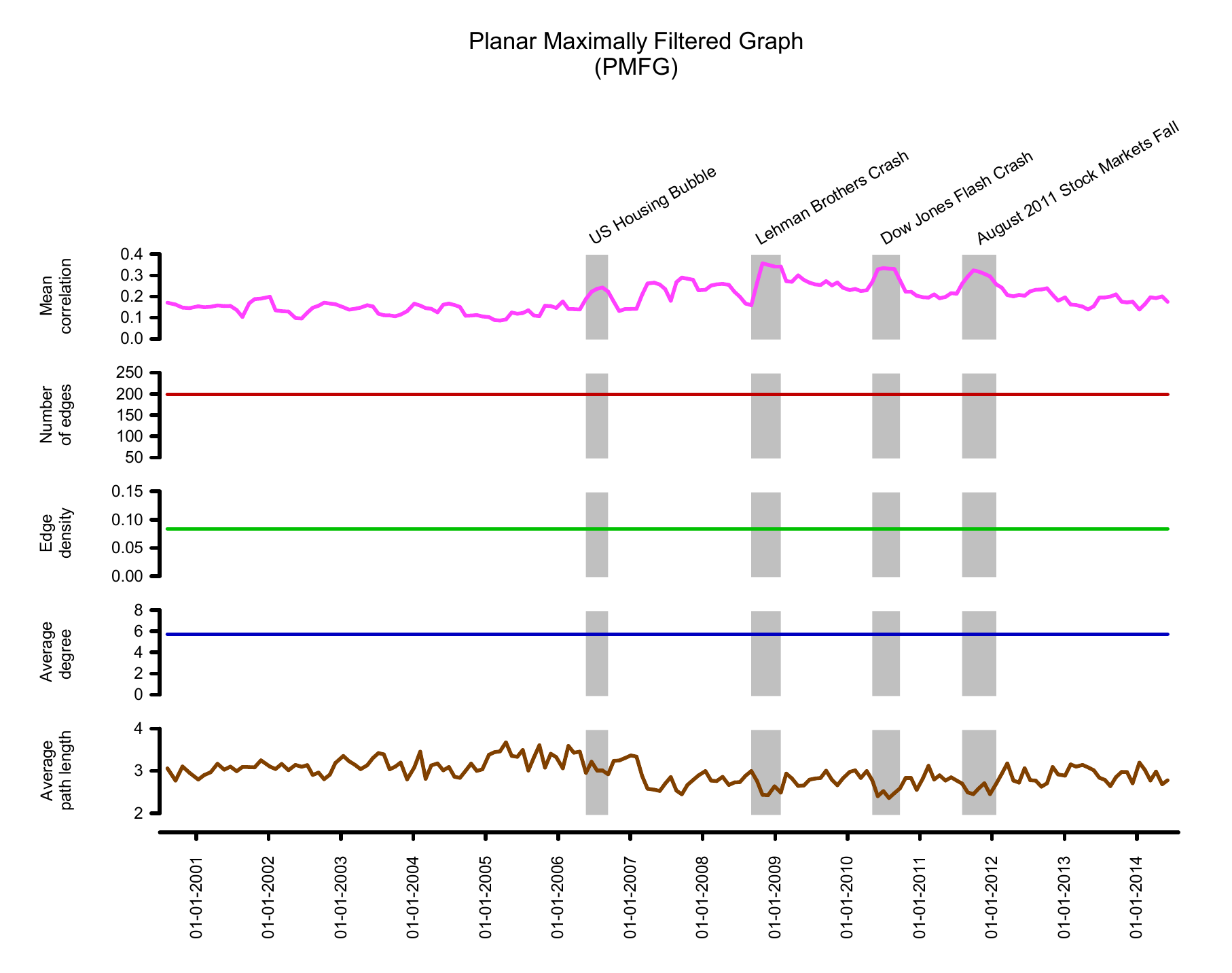}
\end{center}
\caption{Evolution of generic indicators and network characteristics for PMFG 
$\mathbf{P}^{\tau}(t)$ of global market indices, constructed from the 
correlation matrices $\mathbf{C}^{\tau}(t)$ of window size $\tau=80$ days and 
an overlapping shift of $\Delta\tau=20$ days over a period of 14 years 
(2000-2014). From top to bottom, we compare the plot of mean correlation among 
market indices, number of edges, edge density, average degree and average path 
length. The four shaded regions correspond to the epochs around the four important 
market events, namely, US housing bubble, Lehman brothers crash, Dow Jones 
flash crash, and August 2011 stock markets fall.}
\end{figure}
%--------------------------------------------------------------------------

%--------------------------------------------------------------------------
% Figure S12
\begin{figure}[ht]
\label{FigS12}
\begin{center}
\includegraphics[width = 0.72\linewidth]{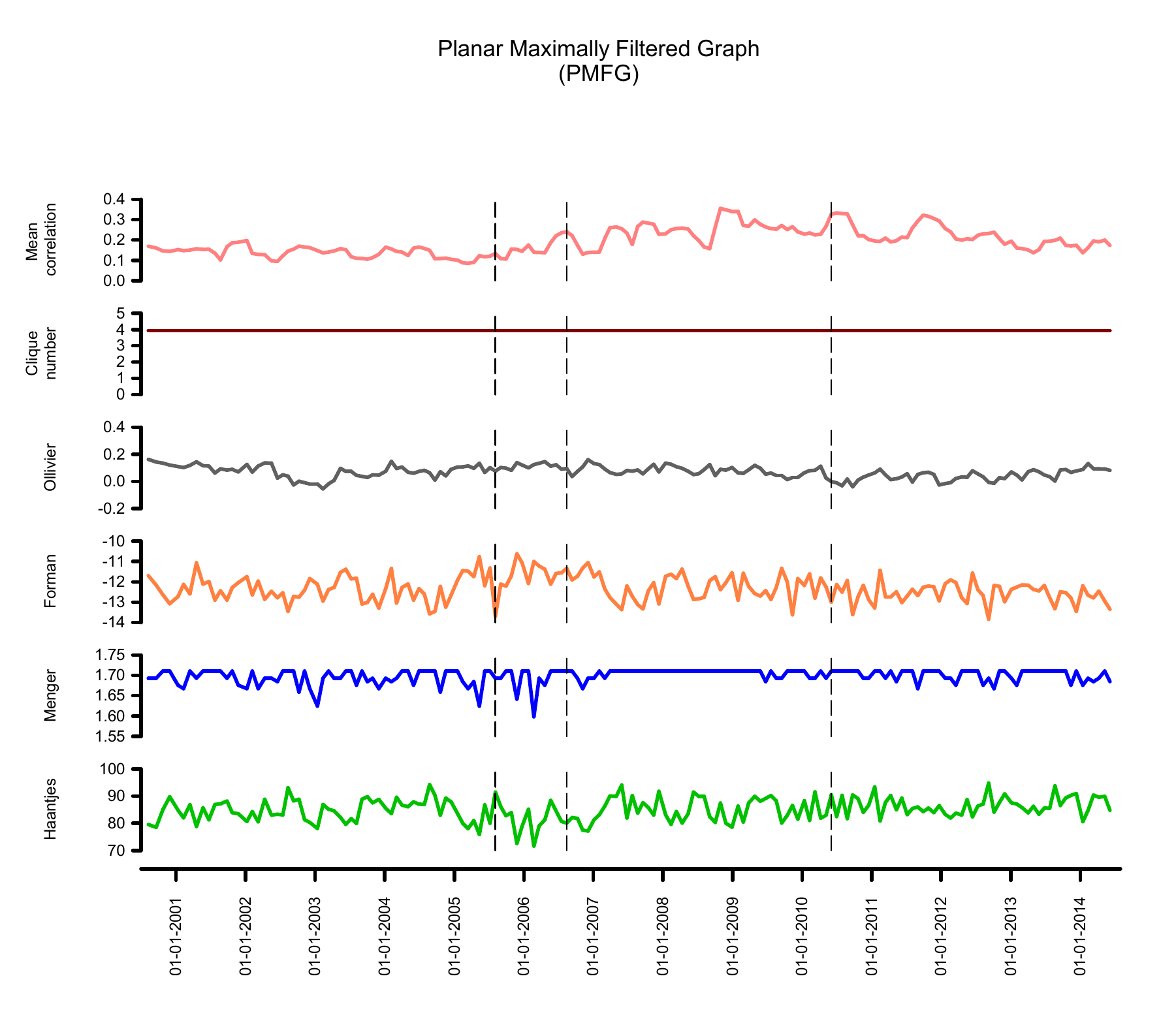}
\end{center}
\caption{Evolution of network characteristics for PMFG $\mathbf{P}^{\tau}(t)$ 
of global market indices with window size $\tau=80$ days and an overlapping 
shift of $\Delta\tau=20$ days. Comparison of the plots of mean correlation 
among market indices, clique number, average of Ollivier-Ricci (OR), 
Forman-Ricci (FR), Menger-Ricci (MR), and Haantjes-Ricci (HR) curvature of 
edges in PMFG over the 14-year period.}
\end{figure}
%--------------------------------------------------------------------------

%--------------------------------------------------------------------------
% Figure S13
\begin{figure}[ht]
\label{FigS13}
\begin{center}
\includegraphics[width = 0.7\linewidth]{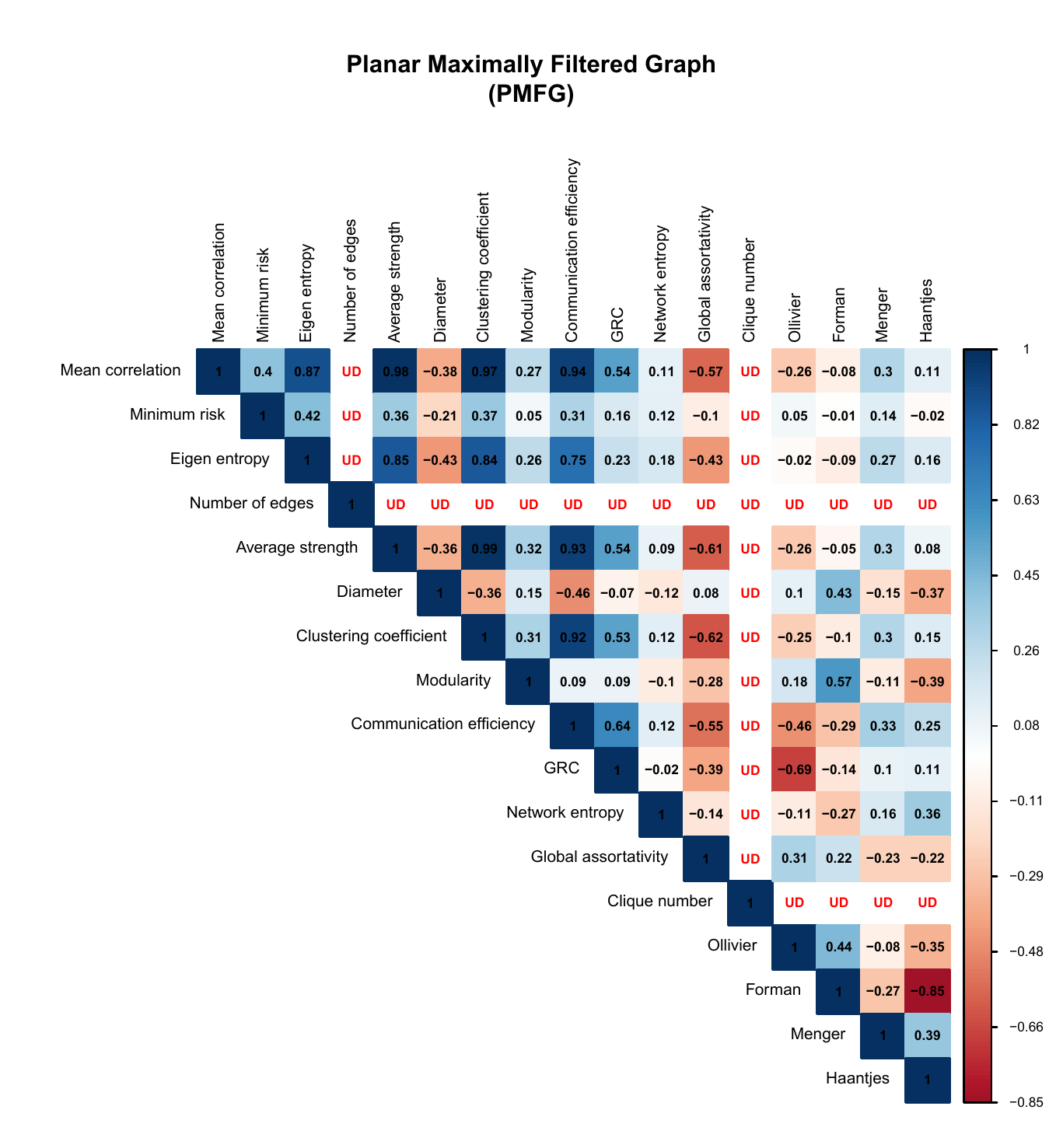}
\end{center}
\caption{Correlation between generic indicators and network characteristics 
for PMFG $\mathbf{P}^{\tau}(t)$ of global market indices, constructed from 
the correlation matrices $\mathbf{C}^{\tau}(t)$ of window size $\tau=80$ 
days and an overlapping shift of $\Delta\tau=20$ days over a period of 14 
years (2000-2014). For some pairs of measures investigated here, the 
correlation cannot be computed as at least one of the measures has zero 
variance, and thus, we specify undefined (`UD') for correlation between 
such pairs of measures in this plot.}
\end{figure}
%--------------------------------------------------------------------------

%--------------------------------------------------------------------------
% Figure S14
\begin{figure}[ht]
\label{FigS14}
\begin{center}
\includegraphics[width = 0.72\linewidth]{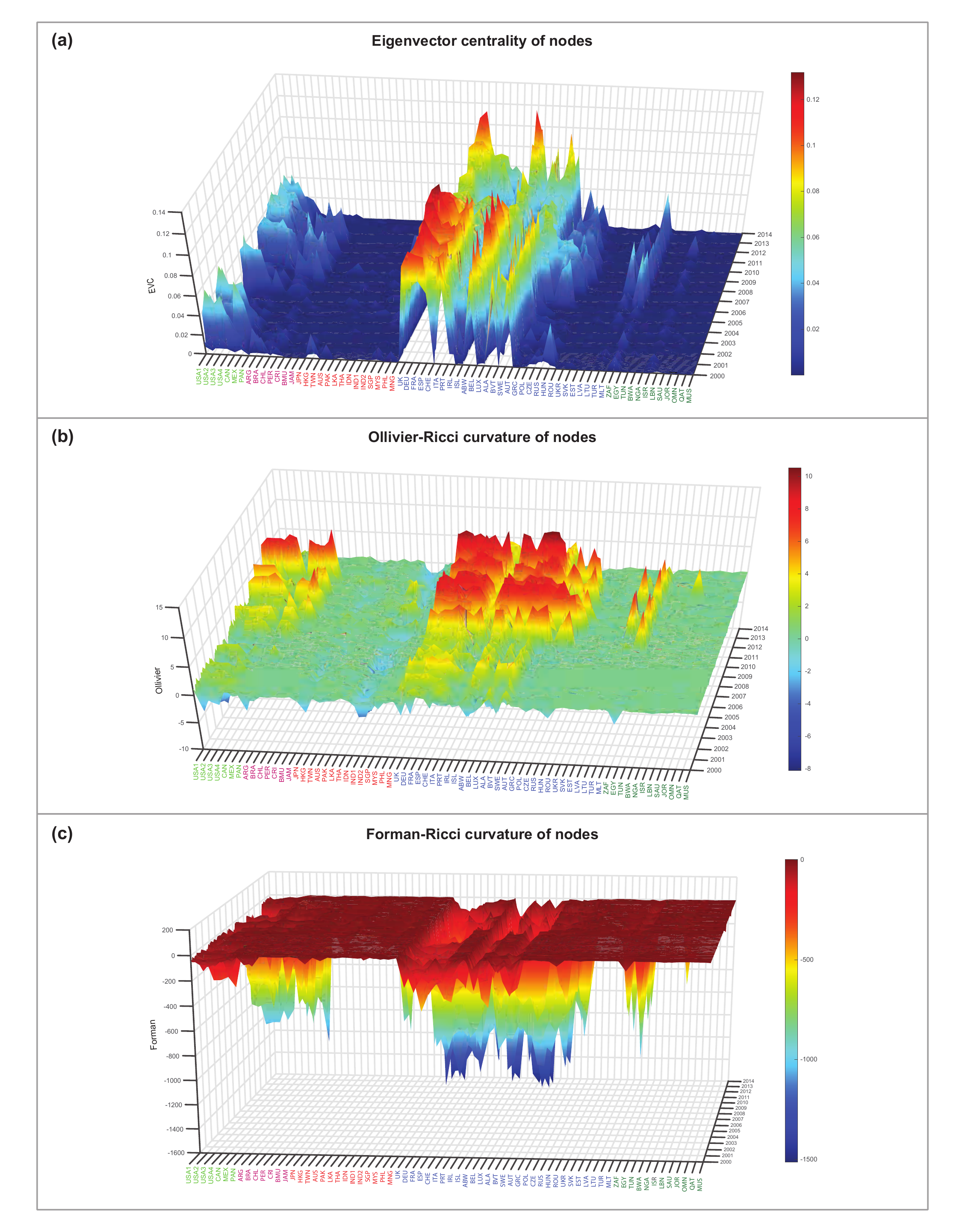}
\end{center}
\caption{Three dimensional plot showing the evolution of {\bf (a)} 
normalized eigenvector centrality, {\bf (b)}  Ollivier-Ricci (OR) 
curvature, and {\bf (c)} Forman-Ricci (FR) curvature for each of 
the 69 market indices (nodes) across the time-series of 172 threshold 
networks $\mathbf{S}^{\tau}(t)$ of market indices computed with window 
size $\tau=80$ days and an overlapping shift of $\Delta\tau=20$ days, 
constructed by adding edges with correlation $C_{ij}^{\tau}(t) \ge 0.65$ 
to the MST. We compute the OR and FR curvature of nodes in the threshold
networks based on the OR and FR curvature of incident edges on each 
node. Here, the colours of the market indices are based on geographical 
region of their country. The four USA market indices, NASDAQ, NYSE, 
RUSSELL1000 and SPX, are labelled as USA1, USA2, USA3 and USA4, 
respectively, while the two Indian indices, NIFTY and SENSEX30, 
are labelled as IND1 and IND2, respectively. It can be seen that there 
exist certain periods of time, when some of the market indices in close 
geographical proximity display high (absolute) values while others 
display low values, indicative of the changes in the complex interactions 
and community structures in global financial network. 
}
\end{figure}
%--------------------------------------------------------------------------

%--------------------------------------------------------------------------
% Supplementary Tables
%--------------------------------------------------------------------------

%--------------------------------------------------------------------------
% Table S1
\begin{table}[ht]
\centering
\caption{List of 69 global financial market indices across 65 countries
considered in this work. For each financial market index, the table
lists the country of origin, country code, index code, region code and
geographical region. This dataset was obtained from Bloomberg.}
\resizebox{\textwidth}{!}{
\begin{tabular}{|p{1cm}|p{5cm}|p{2cm}|p{5cm}|p{2cm}|p{5cm}|}
\hline
\textbf{S. No.} & \textbf{Country}             & \textbf{Country code} & \textbf{Index code}       & \textbf{Region code} & \textbf{Region}                   \\ \hline
1     & United States of America & USA          & NASDAQ COMPOSITE & NA          & North America       \\ \hline
2     & United States of America & USA          & NYSE COMPOSITE   & NA          & North America       \\ \hline
3     & United States of America & USA          & RUSSELL 1000     & NA          & North America       \\ \hline
4     & United States of America & USA          & SPX              & NA          & North America       \\ \hline
5     & Canada                   & CAN          & SPTSX            & NA          & North America       \\ \hline
6     & Mexico                   & MEX          & MEXBOL           & NA          & North America       \\ \hline
7     & Panama                   & PAN          & BVPSBVPS         & NA          & North America       \\ \hline
8     & Argentina                & ARG          & MERVAL           & SA          & South America       \\ \hline
9     & Brazil                   & BRA          & IBOV             & SA          & South America       \\ \hline
10    & Chile                    & CHL          & IPSA             & SA          & South America       \\ \hline
11    & Peru                     & PER          & IGBVL            & SA          & South America       \\ \hline
12    & Costa Rica               & CRI          & BCT              & SA          & South America       \\ \hline
13    & Bermuda                  & BMU          & BSX              & SA          & South America       \\ \hline
14    & Jamaica                  & JAM          & JMSMX            & SA          & South America       \\ \hline
15    & Japan                    & JPN          & TPX              & AP          & Asia Pacific        \\ \hline
16    & Hang Kong                & HKG          & Hang Seng        & AP          & Asia Pacific        \\ \hline
17    & Taiwan                   & TWN          & TWSE             & AP          & Asia Pacific        \\ \hline
18    & Australia                & AUS          & AS51             & AP          & Asia Pacific        \\ \hline
19    & Pakistan                 & PAK          & KSE100           & AP          & Asia Pacific        \\ \hline
20    & Sri Lanka                & LKA          & CSEALL           & AP          & Asia Pacific        \\ \hline
21    & Thailand                 & THA          & SET              & AP          & Asia Pacific        \\ \hline
22    & Indonesia                & IDN          & JCI              & AP          & Asia Pacific        \\ \hline
23    & India                    & IND          & NIFTY            & AP          & Asia Pacific        \\ \hline
24    & India                    & IND          & SENSEX30         & AP          & Asia Pacific        \\ \hline
25    & Singapore                & SGP          & FSSTI            & AP          & Asia Pacific        \\ \hline
26    & Malaysia                 & MYS          & FBMKLCI          & AP          & Asia Pacific        \\ \hline
27    & Philippines              & PHL          & PCOMP            & AP          & Asia Pacific        \\ \hline
28    & Mongolia                 & MNG          & MSETOP           & AP          & Asia Pacific        \\ \hline
29    & United Kingdom           & UK           & UKX              & EME         & Europe Middle East  \\ \hline
30    & Germany                  & DEU          & DAX              & EME         & Europe Middle East  \\  \hline
31    & France                   & FRA          & CAC40            & EME         & Europe Middle East  \\ \hline
32    & Spain                    & ESP          & IBEX35           & EME         & Europe Middle East  \\ \hline
33    & Switzerland              & CHE          & SMI              & EME         & Europe Middle East  \\ \hline
34    & Italy                    & ITA          & FTSEMIB          & EME         & Europe Middle East  \\ \hline
35    & Portugal                 & PRT          & BVLX             & EME         & Europe Middle East  \\ \hline
36    & Ireland                  & IRL          & ISEQ             & EME         & Europe Middle East  \\ \hline
37    & Iceland                  & ISL          & ICEXI            & EME         & Europe Middle East  \\ \hline
38    & Netherlands              & ABW          & AEX              & EME         & Europe Middle East  \\ \hline
39    & Belgium                  & BEL          & BEL20            & EME         & Europe Middle East  \\ \hline
40    & Luxembourg               & LUX          & LUXXX            & EME         & Europe Middle East  \\ 
\hline
41    & Finland                  & ALA          & HEX              & EME         & Europe Middle East  \\ \hline
42    & Norway                   & BVT          & OBX              & EME         & Europe Middle East  \\ \hline
43    & Sweden                   & SWE          & OMX              & EME         & Europe Middle East  \\ \hline
44    & Austria                  & AUT          & ATX              & EME         & Europe Middle East  \\ \hline
45    & Greece                   & GRC          & ASE              & EME         & Europe Middle East  \\ \hline
46    & Poland                   & POL          & WIG              & EME         & Europe Middle East  \\ \hline
47    & Czech Republic           & CZE          & PX               & EME         & Europe Middle East  \\
\hline
48    & Russia                   & RUS          & MICEX            & EME         & Europe Middle East  \\
\hline
49    & Hungary                  & HUN          & BUX              & EME         & Europe Middle East  \\ \hline
50    & Romania                  & ROU          & BET              & EME         & Europe Middle East  \\ \hline
\end{tabular}%
}
\label{indices_table}
\end{table}
%--------------------------------------------------------------------------
\begin{table}[ht]
\centering
\resizebox{\textwidth}{!}{
\begin{tabular}{|p{1cm}|p{5cm}|p{2cm}|p{5cm}|p{2cm}|p{5cm}|}
\hline
51    & Ukraine                  & UKR          & PFTS             & EME         & Europe Middle East  \\ \hline
52    & Slovakia                 & SVK          & SKSM             & EME         & Europe Middle East  \\ \hline
53    & Estonia                  & EST          & TALSE            & EME         & Europe Middle East  \\ 
\hline
54    & Lativa                   & LVA          & RIGSE            & EME         & Europe Middle East  \\ \hline
55    & Lithuania                & LTU          & VILSE            & EME         & Europe Middle East  \\ \hline
56    & Turkey                   & TUR          & XU100            & EME         & Europe Middle East  \\ \hline
57    & Malta                    & MLT          & MALTEX           & EME         & Europe Middle East  \\ \hline
58    & South Africa             & ZAF          & JALSH            & AME         & Africa/Middle East  \\ \hline
59    & Egypt                    & EGY          & HERMES           & AME         & Africa/Middle East  \\ \hline
60    & Tunisia                  & TUN          & TUSISE           & AME         & Africa/Middle East  \\
\hline
61    & Botswana                 & BWA          & BGSMDC           & AME         & Africa/Middle East  \\ \hline
62    & Nigeria                  & NGA          & NGSEINDX         & AME         & Africa/Middle East  \\ \hline
63    & Israel                   & ISR          & TA-25            & AME         & Africa/Middle East  \\ \hline
64    & Lebanon                  & LBN          & BLOM             & AME         & Africa/Middle East  \\ \hline
65    & Saudi Arabia             & SAU          & SASEIDX          & AME         & Africa/Middle East  \\ \hline
66    & Jordan                   & JOR          & JOSMGNFF         & AME         & Africa/Middle East  \\ \hline
67    & Oman                     & OMN          & MSM30            & AME         & Africa/Middle East  \\ \hline
68    & Qatar                    & QAT          & DSM              & AME         & Africa/Middle East  \\ \hline
69    & Mauritius                & MUS          & SEMDEX           & AME         & Africa/Middle East  \\ \hline
\end{tabular}%
}
\end{table}
%--------------------------------------------------------------------------

%--------------------------------------------------------------------------
% Table S2
\begin{table}[ht]
\centering
\caption{Classification of network measures employed to characterize the 
networks of global market indices into those for unweighted or weighted 
graphs. For each measure evaluated in the weighted graph, we list the 
appropriate natural weight, strength or distance, which is used in the 
associated computation.}
{\scriptsize
\begin{tabular}{|l|l|c|c|}
\hline
\textbf{S. No.} & \textbf{Measure}	& \textbf{Type of measure}	& \textbf{Type of weight} \\  
\hline
1				& Number of edges	& Unweighted				&  -				    \\  
\hline
2				& Average degree	& Unweighted 			    &  -					\\  
\hline
3				& Edge density		& Unweighted				&  -					\\  
\hline
4				& Clique number		& Unweighted				&  -					\\  
\hline
5				& Network entropy	& Unweighted				&  -					\\  
\hline
6				& Average weighted degree	& Weighted			& Strength			    \\  
\hline
7				& Global assortativity		& Weighted			& Strength				\\  
\hline
8				& Clustering coefficient	& Weighted			& Strength				\\  
\hline
9				& Modularity				& Weighted			& Strength				\\  
\hline
10				& Eigenvector centrality	& Weighted			& Strength				\\  
\hline
11				& Average shortest path length & Weighted		& Distance				\\	
\hline
12				& Diameter					& Weighted			& Distance				\\  
\hline
13				& Global reaching centrality & Weighted			& Distance				\\  
\hline
14  			& Communication efficiency	& Weighted			& Distance				\\  
\hline
15				& Ollivier-Ricci curvature	& Weighted			& Distance				\\  
\hline
16				& Forman-Ricci curvature	& Weighted			& Distance				\\  
\hline
17				& Menger-Ricci curvature	& Unweighted		& -						\\  
\hline
18			    & Haantjes-Ricci curvature	& Unweighted		& - 					\\  
\hline
\end{tabular}
}
\label{measure_table}
\end{table}
%--------------------------------------------------------------------------

%-----------------------------------------------------------------
\end{document}